%% file: main.tex
\documentclass[numbers]{article}
\usepackage{arxiv}

\input{preamble}

\title{A Comprehensive Survey of Benchmarks for Automated Improvement of Software's Non-Functional Properties}
\renewcommand\shorttitle{A Comprehensive Survey of Benchmarks for Autom. Improvement of Software.'s Non-Func. Properties}
\renewcommand{\undertitle}{}

\author{%
  \href{https://orcid.org/0000-0003-0485-5279}{\includegraphics[scale=0.06]{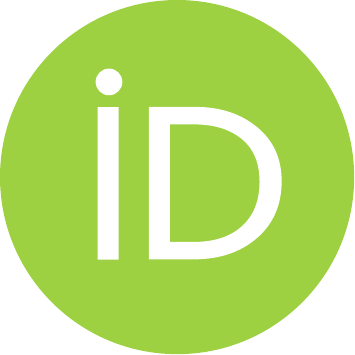}\hspace{1mm}%
    Aymeric~Blot}\\
  Department of Computer Science\\
  University College London, London, U.K.\\[5pt]
  Laboratoire d’Informatique, Signal et Image de la Côte d'Opale\\
  Université du Littoral Côte d'Opale, Calais, France\\
  \href{mailto:aymeric.blot@univ-littoral.fr}{aymeric.blot@univ-littoral.fr}\\
  \And
  \href{https://orcid.org/0000-0002-7833-6044}{\includegraphics[scale=0.06]{orcid.pdf}\hspace{1mm}%
  Justyna~Petke} \\
  Department of Computer Science\\
  University College London, London, U.K.\\
  \href{mailto:j.petke@ucl.ac.uk}{j.petke@ucl.ac.uk}
}

\date{}

\hypersetup{
  pdftitle={A Comprehensive Survey of Benchmarks for Automated Improvement of Software's Non-Functional Properties},
  pdfsubject={cs.SE},
  pdfauthor={Aymeric~Blot, Justyna~Petke},
  pdfkeywords={software performance, non-functional properties, benchmark},
}

\begin{document}

\maketitle

\input{split_abstract}

\input{split_introduction}
\input{split_method}
\input{split_survey}
\input{split_discussion}
\input{split_threats}
\input{split_conclusion}

\bibliographystyle{ACM-Reference-Format}
\bibliography{bib_longest}

\end{document}

%% file: preamble.tex
\usepackage{hyperref}

\usepackage[sort&compress,numbers]{natbib}

\usepackage{amsmath,amsfonts}
\usepackage{graphicx}
\usepackage{textcomp}
\usepackage{xcolor}
\usepackage{tikz}
\usepackage{pgfplots}
\usetikzlibrary{positioning,arrows}
\usepgfplotslibrary{colorbrewer}

\pgfplotsset{
  xlabel near ticks,
  ylabel near ticks,
}

\usepackage{enumitem}
\usepackage{booktabs}
\usepackage{multirow}
\usepackage{algorithm}
\usepackage{listings}
\usepackage[]{algpseudocode}
\algnewcommand{\LineComment}[1]{\State \(\triangleright\) #1}

\usepackage{bm}
\usepackage{placeins}
\usepackage{xspace}

\usepackage{siunitx}
\newcommand{\para}[1]{\smallskip\noindent\textbf{#1}\hskip0.5em\relax}

\newcommand{\cpp}{C\texttt{++}\xspace}
\newcommand{\csharp}{C\texttt{\#}\xspace}

\usepackage{todonotes}

%% file: split_abstract.tex
\begin{abstract}
  Performance is a key quality of modern software.
  Although recent years have seen a spike in research on automated improvement of software's execution time, energy, memory consumption, etc., there is a noticeable lack of standard benchmarks for such work.
  It is also unclear how such benchmarks are representative of current software.
  Furthermore, frequently the non-functional properties of software are targeted for improvement one-at-a-time, neglecting potential negative impact on other properties.

  In order to facilitate more research on automated improvement of non-functional properties of software, we conducted a survey on this topic, gathering benchmarks used in previous work.
  We considered 5 major online repositories of software engineering work: ACM Digital Library, IEEE Xplore, Scopus, Google Scholar, and ArXiV.
  We gathered 5000 publications (3749 unique), which were systematically reviewed to identify work that empirically improves non-functional properties of software.
  We identified 386 relevant papers.

  We find that execution time is the most frequently targeted property for improvement (in 62\% of relevant papers), while multi-objective improvement is rarely considered (5\%).
  Static approaches for automated improvement of non-functional software properties are prevalent (in 53\% of papers), with exploratory approaches (evolutionary in 18\% and non-evolutionary in 14\% of papers) increasingly popular in the last 10 years.
  Only 40\% of 386 papers describe work that uses benchmark suites, rather than single software, of those SPEC is most popular (covered in 33 papers).
  We also provide recommendations for choice of benchmarks in future work, noting, for instance, lack of work that covers Python, JavaScript, mobile devices, and other.
  We provide all programs found in the 386 papers on our dedicated webpage: \url{https://bloa.github.io/nfunc_survey/}.

  We hope that this effort will facilitate more research on the topic of automated improvement of software's non-functional properties.

  \keywords{software performance, non-functional properties, benchmark.}
\end{abstract}

%% file: split_introduction.tex
\section{Introduction}
\label{section:introduction}

The primary focus of software developers is to write bug-free software.
Even then, many issues often arise throughout the development and production cycles, causing significant human resource investment into code maintenance.
Poor software quality is costly.
For example, Krasner~\cite{krasner:2018:cisq} estimated that in 2018 the cost of poor-quality software on the US economy was 2.08 trillion dollars.

In order to deliver better software, many techniques and tools exist to diagnose software's potential flaws, refactor source code, optimise compiled machine code, and even use evolution to automatically derive better software variants.
To this purpose, many surveys have been conducted on techniques for functional improvement of software, particularly regarding automated bug fixing~\cite{legoues:2013:sqj,monperrus:2018:hal_living,monperrus:2018:acmcs,gazzola:2019:tse,cao:2020:isssr}.
We also observed that studies in the field of automated program repair often use well-crafted benchmarks (see, e.g., \url{https://program-repair.org/benchmarks.html}).

On the other hand, software performance in terms of non-functional properties is often relegated to second place.
Execution time, memory usage, energy consumption, for example, are some of such considerations that, despite their importance, are sometimes neglected.
One explanation might relate to the well-known Moore's law, that states that the number of transistors in microchips is doubling every two years.
Indeed, for a very long time, performance was mostly tied to the hardware at hand, and waiting for better hardware was a reasonable performance improvement strategy.
Although Moore's law is still valid in spirit to this day, hardware improvement rate has been slowly decreasing over the past twenty years, making improving software's non-functional properties increasingly compelling.
However, whilst more and more research is now conducted on automated improvement of non-functional properties, the field still lacks standardised benchmarks.
These will help drive the field forward by providing a common baseline for newly proposed approaches.

Therefore, we conducted an in-depth literature review, both to better frame what has been done and is currently conducted, as well as to identify the type of software that is most often targeted.

First, we conduct a preliminary search, hand-picking 100 research articles to identify the most relevant keywords linked to automated improvement of non-functional properties of software.
We then query five major online repositories for related work---ACM Digital Library, IEEE Xplore, Scopus, Google Scholar, and ArXiV---grouping useful keywords into five subsets and considering the first 200 results each time.
Out of 5000 results returned we found 3749 unique papers which we systematically checked for empirical work that improves non-functional properties of software, providing potential benchmarks for future work.
Combined with the work picked in the preliminary search, this gave us 386 publications, that we then categorised with regards to the property they target, the type of approach they use, and central to our survey, the benchmark they consider.

With our survey, we aim to answer the following research questions.

\begin{description}
\item[RQ1 (Existing Benchmarks)]
  How prevalent is empirical work on automated non-functional performance improvement of software?
\item[\quad (a)] What type of non-functional property is most often improved?
\item[\quad (b)] When optimised together, which combinations of non-functional properties are considered?
\item[\quad (c)] Which approaches for non-functional improvement are used most often?
\item[\quad (d)] How is software most often modified?

\item[RQ2 (Targeted Software)]
  What software is used to validate work on improvement of non-functional properties of software?
\item[\quad (a)] How often are existing software benchmarks reused?
\item[\quad (b)] Which software is targeted most often for improvement?

\item[RQ3 (Software Diversity)]
  How representative are the benchmarks used in work on non-functional improvement of real-world software?
\item[\quad (a)] What type of software is targeted most frequently?
\item[\quad (b)] Which programming languages are targeted most frequently?
\end{description}

The main insights from our survey are:
\begin{itemize}
\item There is a large amount of literature pertaining to improving non-functional properties of software, starting around 25 years ago, becoming increasingly popular in recent years.
\item Despite the large amount of work, there are very few prominent software benchmarks.
\item Finally, we found clear discrepancies between the software characteristics from academic work and those reported by industry and online surveys.
\end{itemize}

All surveyed work, including benchmarks used, is available online\footnote{\url{https://bloa.github.io/nfunc_survey}}.
We hope this resource will facilitate more work on automated improvement of non-functional properties of software.

%% file: split_method.tex
\section{Survey Methodology}
\label{section:method}

We conducted a literature review in order to establish state-of-the-art in automated improvement of software's non-functional properties.
We started with a preliminary search, to construct a set of keywords for a subsequent, larger search over five repositories.
These two steps are described next.

\subsection{Preliminary Search}

Whilst strongly anchored in the software engineering world, work on software optimisation spans many different independent research fields that do not necessarily use a consistent terminology, let alone share a unified one.
In order to conduct an adequate literature review, we first needed to make sure relevant keywords were used.
To that purpose, we performed a manual search, gathering a small number of diverse set of relevant work on improvement of software's non-functional properties.
We use a simple criterion to establish whether a piece of work qualifies as improving a non-functional property of software, namely if the intended semantics of the transformed software is preserved.
In that sense work on automated optimisation of runtime, energy, or memory consumption is deemed relevant, whilst work on bug fixing or software transplantation is not relevant, as, by definition, the input/output behaviour of the transformed software will change.
The only exception to this rule is software specialisation, where functionality could be compromised for improvement of a non-functional property.
Since the main focus of such work is improvement of non-functional behaviour, we still consider it relevant.

Starting from known related work, we iteratively built a purposely diverse corpus of 100 relevant publications by querying specific research fields (e.g., ``genetic improvement'', ``code refactoring'', ``compiler tuning''), specific types of non-functional properties (e.g., ``reliability'', ``complexity''), using synonyms (e.g., ``software evolution'' and ``program evolution'', ``energy consumption'' and ``energy footprint''), etc.
We then extracted, from the title and abstract of every selected publication, every word that could potentially be used as a keyword during the systematic repository search.
We calculated how frequently each word occurs in the metadata of selected work.
In addition, we investigated the use of wildcards, grouping words sharing similar prefixes, and expressions (e.g., ``execution time'' or ``running time'', as opposed to ``time'' alone).
We also tried to singularise words, removing final ``s''-es when the pluralised versions of words were also found.

\input{figure/keywords}

\input{table/keywords}
Details on the frequency analysis are given in \autoref{figure:keywords}.
Note that we excluded prepositions, articles, and other generic words clearly not useful for our literature survey (e.g., ``result'', ``paper'', ``is'', ``are'').
Surprisingly, word combinations did not result in particularly frequent expressions, appearing significantly less often than their respective individual words.
However, prefix wildcards were very effective in providing usable search keywords.
Then, a subset of the most frequent words was derived and further classified into seven classes of potential keywords, as detailed in \autoref{table:keywords}.
These keywords were used as a basis for the second step of our survey protocol.

\subsection{Systematic Repository Search}

Based on keywords found in our preliminary search, we perform systematic search for relevant work, largely inspired by the methodology proposed by Hort~et~al.~\cite{hort:2022:tse}.
We use three groups of keywords: two that ensure papers relate to software performance improvement (\emph{Software} and \emph{Improvement} in \autoref{table:keywords}) and one group of keywords targeting a specific non-functional property (including \emph{Time}, \emph{Memory}, \emph{Energy}, \emph{Quality}, and \emph{Others}).
The addition of the supplementary \emph{Improvement} group of keywords is motivated by the large number of papers otherwise returned by each search.
In total, five separate queries are therefore constructed, one for each keyword from the \emph{Non-Functional Property} category.
When applied on the preliminary dataset of 100 papers, they collectively achieve 97\% coverage, with only one paper missing a \emph{Software} keyword (\cite{moyasebi:2020:gi-gecco}) and two missing an \emph{Improvement} keyword (\cite{poor:2020:erlang-icfp,duan:2020:secdev}).

These five queries were used, to ensure good representation, on five major digital libraries: ACM Digital Library, IEEE Xplore, Scopus, Google Scholar, and ArXiV, for a total of 25 searches.
In particular, Springer Link, Science Direct, or JSTOR were not considered due to their inability to handle complex Boolean queries or queries with many keywords.
Where filters allowed for it, each search was restricted to the computer science research field (Scopus, ArXiV), as well as restricted to conferences proceedings and journal articles (ACM, IEEE, Scopus).
We made no further restrictions.
Because of the high number of papers returned by the 25 queries, we only focus on the first 200 papers returned by the digital libraries, using the provided default relevance-based sort order.

Overall, 5,000 papers were considered, and after deduplication, the following criteria were used for selection, which define the scope of our survey:

\para{Inclusion Criteria}
A paper is deemed relevant when it fulfils the following four criteria:
\begin{enumerate}
\item it must relate to a quantifiable non-functional property;
\item it must contain an empirical study which applies a software improvement technique to existing software; 
\item improvement of the targeted non-functional property must be the active focus of the paper and not merely a side-effect;
\item the approach used must result in a distinct software execution that can thus be compared to the original software.
\end{enumerate}

\para{Selection Process}
Publications are then categorised according to a three-step selection process:
\begin{description}
\item[Title:] first, publications whose titles clearly do not fit the spirit of the survey are discarded without further reading;
\item[Abstract:] second, abstracts are inspected and publications are rejected when at least one inclusion criteria clearly does not apply;
\item[Body:] only then remaining potentially relevant publications are read in full and included, depending on the relevance of their content.
\end{description}

\input{table/systematic_pass.tex}
Details on the systematic repository search are given in \autoref{table:systematic_pass}.
In particular, for each class of query and each online digital library we present the total number of hits, and the numbers of papers rejected at each step of the selection process, or ultimately classified as relevant.

In total, the 25 queries yielded 5000 results.
Papers with identical title were merged together after manual verification, resulting in overall 3749 unique papers (25\% redundancy between query terms and online repositories).
Note that in some rare cases individual queries resulted in fewer than 200 unique papers; this is, for example, due to some papers having multiple DOIs\footnote{E.g., a paper published in OOPSLA 2010 (\url{https://doi.org/10.1145/1869459.1869473}) also published in ACM SIGPLAN Notices (\url{https://doi.org/10.1145/1932682.1869473}), or a paper published in ICCAD 2006 differently indexed by IEEE (\url{https://doi.org/10.1109/ICCAD.2006.320144}) and ACM (\url{https://doi.org/10.1145/1233501.1233551})}.

Of these 3749 unique papers, we determined 2151 to be irrelevant based on their title alone (e.g., ``Memory training and memory improvement in Alzheimer's disease: Rules and exceptions'' is clearly not relevant), 136 to be irrelevant based on their abstract, and finally 70 to be irrelevant based on their actual content.
Overall, this second step of the survey yielded 305 unique relevant papers (8.14\%).
This manual step was conducted over the span of two months. 

We observed that Scopus yielded by far the highest number of relevant papers, almost four times the number from Google Scholar and ArXiV combined.
All types of queries yielded similar numbers of papers, with the exception of the ``Time'' keyword group, although \autoref{section:survey} will show that execution time is by far the most common non-functional property optimised in the literature.

\begin{figure}
  \begin{minipage}[t]{0.48\textwidth}
    \input{figure/postmortem_lib.tex}
  \end{minipage}\hfill
  \begin{minipage}[t]{0.48\textwidth}
    \input{figure/venn.tex}
  \end{minipage}
\end{figure}

Finally, as means to validate the threshold of 200 papers investigated for every query, we investigated the rate at which relevant work appears throughout the systematic repository search.
\autoref{figure:post:lib} shows, for each of the five digital libraries, how many would have been returned had a smaller threshold been used.
Very surprisingly, the rate according to which relevant work is found is almost constant, meaning that a considerable amount of relevant work can be expected to be found even after our threshold of 200 papers per query.
Similar rates are also observed when controlling for each of the keywords categories, i.e., there is a lot more related work for all types of non-functional properties.

\subsection{Corpus}

\input{table/papers_summary.tex}

In preliminary search we identified 100 relevant papers.
The systematic repository search yielded 304 unique papers.
Combined, they resulted in 386 unique relevant papers on the topic of improvement of non-functional properties of software, as shown in \autoref{table:relevant_papers}.
We note that despite a theoretical 97\% coverage, only 19 of the 100 papers from the preliminary search were actually rediscovered during the systematic repository search.
This can be explained by the large number of hits returned by every query (see \autoref{table:systematic_pass}) and the consistent rate at which relevant work is found (see \autoref{figure:post:lib}).
Recall that in the systematic search we only considered relevant work located in the first 200 results of the 25 queries of the repository search.
On the one hand, this means that only 19\% of our hand-picked papers appear within that threshold, once again corroborating the idea that many more relevant pieces of work exist in the literature.
However, on the other hand, it also means that most of the relevant work identified in the systematic search is new, reducing potential unconscious bias from the preliminary search.

By combining both the preliminary and the systematic repository search we thus construct a very rich and diverse corpus of 386 relevant publications, that is, by construction, both relevant in terms of coverage of the different aspects of non-functional improvement, as well as in terms of statistical representativity, as much as this is possible using digital libraries.
In what follows, with the only exception of \autoref{figure:repo_by_fitness} and \autoref{figure:repo_by_fitness2} that only use the 304 repository papers, all discussions and figures are based on the full corpus of 386 papers.

\autoref{figure:venn} presents in more detail the origin of the 386 corpus papers.
Both sets originating from Google Scholar (26 papers) and ArXiV (10 papers) are very small and almost disconnected from the other sets of work.
In fact, the vast majority of publications (85\%) is only found once, and with the exception of Scopus covering a non-negligible number of publications from ACM (20) and IEEE (26).
Despite this, the corpus contains two papers simultaneously returned by ACM, IEEE, Scopus, and our manual search~\cite{xin:2020:ase,sharif:2018:ase}.

%% file: figure/keywords.tex
\begin{figure}
  \centering
  \begin{tikzpicture}
    \begin{axis}[
        width=30em,
        xbar,
        xmin=0,
        xlabel={Number of papers matching keyword},
        ylabel={Keyword found in title or abstract},
        ytick={0,1,2,3,4,5,6,7,8,9,10,11,12,13,14,15,16,17,18,19,20,21,22,23,24,25,26,27,28,29,30,31,32,33,34,35,36,37,38,39,40,41,42,43,44,45,46,47,48,49,50,51,52,53,54,55,56,57,58,59,60,61,62,63,64,65,66},
        yticklabels={structure,setting,improving,give,functional,faster,configuration,complexity,variant,quality,evolve,efficiency,refactoring,parameter,evolution,accura*,energy consumption,space,optimizing,increase,evolutionary,efficient,effective,performance improvement,size,reducing,functionality,energy,speed,runtime,fast*,source code,transformation,optimize,consumption,increas*,speed*,tuning,automatically,automated,effective*,genetic programming,reduce,transform*,functional*,effic*,source,compil*,genetic improvement,optimization,evol*,programming,reduc*,search,automat*,improve,application,genetic,time,software,program,performance,improvement,optim*,code,improv*},
        yticklabel style={font=\small},
        nodes near coords={\pgfkeys{/pgf/fpu}\pgfmathparse{\pgfplotspointmeta*100/100}\pgfmathprintnumber[fixed,precision=0]{\pgfmathresult}\,\%},
        nodes near coords align={horizontal},
        bar shift=0pt,
        enlarge x limits={upper, value=0.20},
        enlarge y limits={true, abs value=0.80},
        bar width=0.6em,
        y=0.8em,
        legend pos={south east},
        thick,
      ]
      \addplot coordinates {
        (10,0) 
        (10,1) 
        (10,2) 
        (10,3) 
        (10,4) 
        (10,5) 
        (10,6) 
        (10,7) 
        (11,8) 
        (11,9) 
        (11,10) 
        (11,11) 
        (12,12) 
        (12,13) 
        (12,14) 
        (13,17) 
        (13,18) 
        (13,19) 
        (13,20) 
        (13,21) 
        (13,22) 
        (14,24) 
        (14,25) 
        (14,26) 
        (15,27) 
        (16,28) 
        (17,29) 
        (19,32) 
        (19,33) 
        (19,34) 
        (21,37) 
        (21,38) 
        (22,39) 
        (23,42) 
        (28,46) 
        (30,49) 
        (33,51) 
        (36,53) 
        (42,55) 
        (42,56) 
        (47,57) 
        (50,58) 
        (52,59) 
        (54,60) 
        (54,61) 
        (54,62) 
        (67,64) 
      };
      \addplot coordinates {
        (12,15) 
        (18,30) 
        (20,35) 
        (20,36) 
        (22,40) 
        (24,43) 
        (25,44) 
        (26,45) 
        (28,47) 
        (32,50) 
        (35,52) 
        (39,54) 
        (57,63) 
        (69,65) 
      };
      \addplot coordinates {
        (12,16) 
        (13,23) 
        (18,31) 
        (22,41) 
        (28,48) 
      };
      \legend{Single keyword,Wildcard keyword,Keyword combination};
    \end{axis}
  \end{tikzpicture}
  \caption{Preliminary search: frequent words in titles and abstracts. ($\ge$10\%)}
  \label{figure:keywords}
\end{figure}
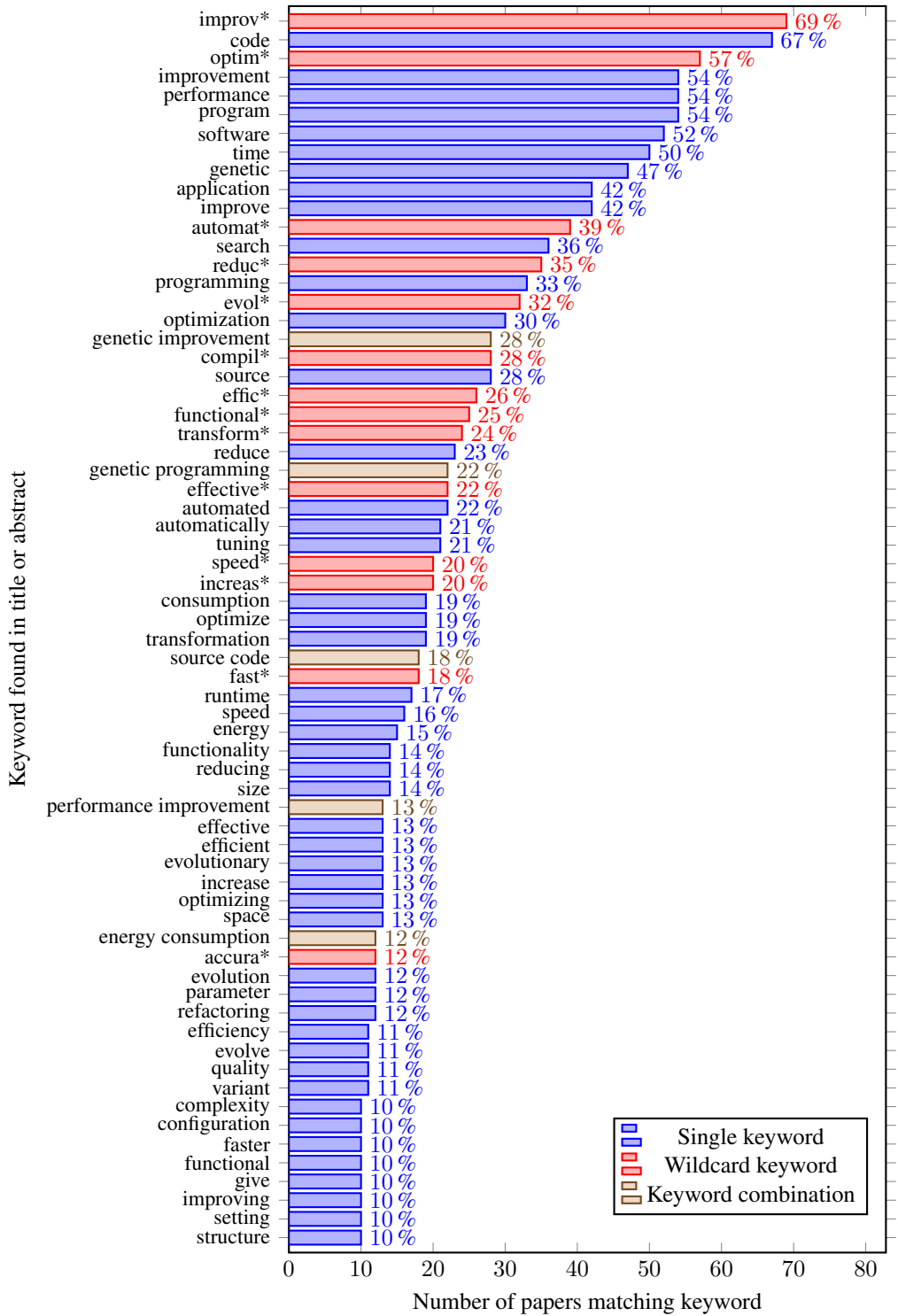

%% file: table/keywords.tex
\begin{table}[t!]
  \centering
  \caption{Keywords used in the systematic repository search.}
  \label{table:keywords}
  \begin{tabular}{llp{25em}}
    \toprule
    \textbf{Group} & \textbf{Category} & \textbf{Keywords}\\
    \midrule
    Software &  & code, program, software, application\\
    \cmidrule(lr){1-3}
    Improvement &    & optim* (optimize, optimizing, optimization), improv* (improve, improving, improvement), automat* (automated, automatically), reduc* (reduce, reducing)\\
    \cmidrule(lr){1-3}
    Non-Functional & Time     & time, runtime, speed* (speed, speedup), fast* (fast, faster)\\
    Property & Memory   & memory\\
    & Energy   & energy, power\\
    & Quality  & performance, effic* (efficient, efficiency), effective* (effective, effectiveness), accura* (accuracy), precis* (precision)\\
    & Other   & functional* (functional, functionality), size, slim* (slimming), bloat, debloating\\
    \bottomrule
  \end{tabular}

  \medskip
  Wildcards (``*'') are used on digital libraries supporting such queries; otherwise we list the alternative keywords used.
\end{table}

%% file: table/systematic_pass.tex
\begin{table}[t!]
  \caption{Systematic repository search.
    For each of the 25 queries we detail how many papers were returned, how many unique ones are found within the first 200 hits, and how many were considered after each of the three steps of our selection process.
    We also indicate for every step how many unique papers are found, both across all five digital libraries (in the last column) as well as across all five types of queries (last rows).}
  \centering
  \begin{tabular}{clcccccc}
    \toprule
    & Step & ACM & IEEE & Scopus & Scholar & ArXiV & Unique\\
    \midrule
    \parbox[t]{2mm}{\multirow{5}{*}{\rotatebox[origin=c]{90}{Time}}}
    & Total hits & 26K & 216K & 375K & 6.3M & 1268 & --\\
    & Unique (first 200 hits) & 200 & 199 & 200 & 200 & 200 & 998\\
    & Selected (Title)        & 60 & 84 & 16 & 55 & 31 & 245\\
    & Selected (Abstract)     & 39 & 52 & 3 & 18 & 13 & 124\\
    & Selected (Body)         & 16 & 29 & 1 & 7 & 2 & 55\\

    \cmidrule(lr){1-8}
    \parbox[t]{2mm}{\multirow{5}{*}{\rotatebox[origin=c]{90}{Memory}}}
    & Total hits              & 6326 & 30K & 57K & 5.2M & 273 & --\\
    & Unique (first 200 hits) & 200 & 199 & 198 & 200 & 199 & 947\\
    & Selected (Title)        & 78 & 88 & 84 & 66 & 56 & 347\\
    & Selected (Abstract)     & 42 & 61 & 62 & 33 & 28 & 209\\
    & Selected (Body)         & 18 & 37 & 32 & 13 & 5 & 96\\

    \cmidrule(lr){1-8}
    \parbox[t]{2mm}{\multirow{5}{*}{\rotatebox[origin=c]{90}{Energy}}}
    & Total hits              & 7907 & 140K & 154K & 5.1M & 727 & --\\
    & Unique (first 200 hits) & 199 & 200 & 198 & 200 & 200 & 962\\
    & Selected (Title)        & 93 & 80 & 111 & 47 & 41 & 350\\
    & Selected (Abstract)     & 56 & 46 & 84 & 8 & 18 & 194\\
    & Selected (Body)         & 16 & 27 & 41 & 1 & 2 & 75\\

    \cmidrule(lr){1-8}
    \parbox[t]{2mm}{\multirow{5}{*}{\rotatebox[origin=c]{90}{Quality}}}
    & Total hits              & 44K & 351K & 591K & 5.7M & 1439 & --\\
    & Unique (first 200 hits) & 200 & 200 & 199 & 200 & 200 & 974\\
    & Selected (Title)        & 97 & 79 & 94 & 43 & 36 & 333\\
    & Selected (Abstract)     & 55 & 50 & 71 & 15 & 14 & 193\\
    & Selected (Body)         & 26 & 34 & 31 & 6 & 2 & 91\\

    \cmidrule(lr){1-8}
    \parbox[t]{2mm}{\multirow{5}{*}{\rotatebox[origin=c]{90}{Others}}}
    & Total hits              & 8166 & 63K & 116K & 6.1M & 826 & --\\
    & Unique (first 200 hits) & 199 & 198 & 192 & 200 & 199 & 938\\
    & Selected (Title)        & 93 & 75 & 99 & 39 & 35 & 308\\
    & Selected (Abstract)     & 67 & 47 & 82 & 14 & 15 & 195\\
    & Selected (Body)         & 36 & 32 & 56 & 5 & 2 & 105\\

    \midrule
    \midrule
    \parbox[t]{2mm}{\multirow{4}{*}{\rotatebox[origin=c]{90}{Overall}}}
    & Unique (first 200 hits) & 881 & 738 & 887 & 844 & 555 & 3749\\
    & Selected (Title)        & 356 & 278 & 334 & 204 & 109 & 1188\\
    & Selected (Abstract)     & 214 & 172 & 242 & 74 & 49 & 678\\
    & Selected (Body)         & 87 & 105 & 126 & 26 & 10 & 305\\

    \bottomrule
    \label{table:systematic_pass}
  \end{tabular}
\end{table}

%% file: figure/postmortem_lib.tex
  \centering
  \begin{tikzpicture}[]
    \begin{axis}[
        width=20em,
        height=16em,
        ymin=0,
        xmin=1,
        xmax=1300,
        xlabel={\small Depth of the repository search},
        ylabel={\small Unique relevant papers},
        xticklabel={\pgfmathparse{\tick/5}\pgfmathprintnumber{\pgfmathresult} hits},
        xticklabel style={rotate=90},
        xtick={200,400,600,800,1000},
        grid=both,
        grid style={line width=.1pt, draw=gray!40},
        major grid style={line width=.2pt,draw=gray!75},
        minor tick num=1,
        no markers,
        thick,
      ]
      \addplot+[red] table[x=x,y=y,meta=l]{
x y l
5 1 acm
10 2 acm
15 3 acm
20 3 acm
25 4 acm
30 4 acm
35 6 acm
40 9 acm
45 10 acm
50 10 acm
55 11 acm
60 11 acm
65 13 acm
70 15 acm
75 15 acm
80 16 acm
85 16 acm
90 17 acm
95 19 acm
100 20 acm
105 20 acm
110 20 acm
115 20 acm
120 21 acm
125 21 acm
130 21 acm
135 23 acm
140 23 acm
145 23 acm
150 23 acm
155 23 acm
160 24 acm
165 25 acm
170 25 acm
175 25 acm
180 26 acm
185 26 acm
190 26 acm
195 27 acm
200 29 acm
205 29 acm
210 29 acm
215 29 acm
220 30 acm
225 30 acm
230 30 acm
235 32 acm
240 32 acm
245 32 acm
250 33 acm
255 35 acm
260 36 acm
265 36 acm
270 36 acm
275 36 acm
280 36 acm
285 37 acm
290 38 acm
295 38 acm
300 38 acm
305 40 acm
310 40 acm
315 40 acm
320 41 acm
325 41 acm
330 42 acm
335 43 acm
340 43 acm
345 44 acm
350 45 acm
355 45 acm
360 45 acm
365 45 acm
370 46 acm
375 46 acm
380 46 acm
385 47 acm
390 49 acm
395 50 acm
400 50 acm
405 50 acm
410 51 acm
415 52 acm
420 52 acm
425 54 acm
430 55 acm
435 55 acm
440 55 acm
445 55 acm
450 56 acm
455 56 acm
460 58 acm
465 59 acm
470 59 acm
475 60 acm
480 60 acm
485 60 acm
490 60 acm
495 60 acm
500 60 acm
505 61 acm
510 61 acm
515 61 acm
520 62 acm
525 62 acm
530 62 acm
535 62 acm
540 62 acm
545 62 acm
550 62 acm
555 63 acm
560 63 acm
565 63 acm
570 64 acm
575 65 acm
580 65 acm
585 65 acm
590 65 acm
595 66 acm
600 66 acm
605 66 acm
610 66 acm
615 66 acm
620 66 acm
625 66 acm
630 67 acm
635 67 acm
640 67 acm
645 67 acm
650 67 acm
655 67 acm
660 68 acm
665 68 acm
670 68 acm
675 70 acm
680 70 acm
685 71 acm
690 71 acm
695 72 acm
700 72 acm
705 72 acm
710 72 acm
715 74 acm
720 74 acm
725 75 acm
730 75 acm
735 75 acm
740 75 acm
745 75 acm
750 75 acm
755 76 acm
760 76 acm
765 76 acm
770 76 acm
775 76 acm
780 76 acm
785 76 acm
790 77 acm
795 77 acm
800 77 acm
805 77 acm
810 77 acm
815 77 acm
820 78 acm
825 78 acm
830 79 acm
835 79 acm
840 79 acm
845 79 acm
850 79 acm
855 80 acm
860 81 acm
865 82 acm
870 84 acm
875 84 acm
880 84 acm
885 84 acm
890 84 acm
895 84 acm
900 84 acm
905 84 acm
910 84 acm
915 84 acm
920 84 acm
925 84 acm
930 84 acm
935 84 acm
940 86 acm
945 86 acm
950 87 acm
955 87 acm
960 87 acm
965 87 acm
970 87 acm
975 87 acm
980 87 acm
985 87 acm
990 87 acm
995 87 acm
1000 87 acm
} node[right] {ACM};
      \addplot+[blue] table[x=x,y=y,meta=l]{
x y l
5 1 ieee
10 3 ieee
15 5 ieee
20 7 ieee
25 8 ieee
30 9 ieee
35 9 ieee
40 11 ieee
45 12 ieee
50 13 ieee
55 14 ieee
60 14 ieee
65 14 ieee
70 15 ieee
75 16 ieee
80 18 ieee
85 19 ieee
90 19 ieee
95 19 ieee
100 20 ieee
105 20 ieee
110 20 ieee
115 20 ieee
120 20 ieee
125 20 ieee
130 20 ieee
135 20 ieee
140 21 ieee
145 22 ieee
150 22 ieee
155 24 ieee
160 25 ieee
165 26 ieee
170 26 ieee
175 27 ieee
180 27 ieee
185 29 ieee
190 29 ieee
195 30 ieee
200 31 ieee
205 31 ieee
210 31 ieee
215 31 ieee
220 32 ieee
225 33 ieee
230 35 ieee
235 35 ieee
240 36 ieee
245 36 ieee
250 37 ieee
255 38 ieee
260 39 ieee
265 39 ieee
270 39 ieee
275 41 ieee
280 41 ieee
285 41 ieee
290 41 ieee
295 41 ieee
300 41 ieee
305 41 ieee
310 42 ieee
315 42 ieee
320 44 ieee
325 44 ieee
330 45 ieee
335 45 ieee
340 45 ieee
345 47 ieee
350 47 ieee
355 49 ieee
360 49 ieee
365 49 ieee
370 50 ieee
375 50 ieee
380 51 ieee
385 51 ieee
390 52 ieee
395 52 ieee
400 53 ieee
405 54 ieee
410 56 ieee
415 57 ieee
420 57 ieee
425 57 ieee
430 57 ieee
435 57 ieee
440 58 ieee
445 58 ieee
450 58 ieee
455 58 ieee
460 58 ieee
465 58 ieee
470 59 ieee
475 59 ieee
480 59 ieee
485 60 ieee
490 60 ieee
495 60 ieee
500 60 ieee
505 60 ieee
510 60 ieee
515 60 ieee
520 60 ieee
525 60 ieee
530 60 ieee
535 60 ieee
540 61 ieee
545 62 ieee
550 62 ieee
555 62 ieee
560 63 ieee
565 63 ieee
570 64 ieee
575 65 ieee
580 65 ieee
585 65 ieee
590 66 ieee
595 66 ieee
600 67 ieee
605 67 ieee
610 68 ieee
615 68 ieee
620 69 ieee
625 69 ieee
630 70 ieee
635 70 ieee
640 71 ieee
645 72 ieee
650 73 ieee
655 73 ieee
660 74 ieee
665 75 ieee
670 75 ieee
675 75 ieee
680 75 ieee
685 76 ieee
690 76 ieee
695 76 ieee
700 76 ieee
705 76 ieee
710 77 ieee
715 77 ieee
720 78 ieee
725 78 ieee
730 79 ieee
735 79 ieee
740 79 ieee
745 80 ieee
750 80 ieee
755 81 ieee
760 81 ieee
765 81 ieee
770 81 ieee
775 82 ieee
780 83 ieee
785 83 ieee
790 83 ieee
795 83 ieee
800 83 ieee
805 84 ieee
810 85 ieee
815 85 ieee
820 85 ieee
825 85 ieee
830 87 ieee
835 87 ieee
840 87 ieee
845 87 ieee
850 88 ieee
855 88 ieee
860 88 ieee
865 89 ieee
870 89 ieee
875 91 ieee
880 92 ieee
885 93 ieee
890 93 ieee
895 94 ieee
900 95 ieee
905 95 ieee
910 96 ieee
915 97 ieee
920 98 ieee
925 99 ieee
930 100 ieee
935 100 ieee
940 100 ieee
945 100 ieee
950 101 ieee
955 103 ieee
960 103 ieee
965 103 ieee
970 103 ieee
975 103 ieee
980 104 ieee
985 105 ieee
990 105 ieee
995 105 ieee
1000 105 ieee
} node[right] {IEEE};
      \addplot+[green!50!black] table[x=x,y=y,meta=l]{
x y l
5 3 scopus
10 4 scopus
15 6 scopus
20 7 scopus
25 9 scopus
30 10 scopus
35 11 scopus
40 12 scopus
45 14 scopus
50 17 scopus
55 18 scopus
60 19 scopus
65 19 scopus
70 22 scopus
75 22 scopus
80 23 scopus
85 25 scopus
90 25 scopus
95 25 scopus
100 27 scopus
105 28 scopus
110 29 scopus
115 31 scopus
120 33 scopus
125 34 scopus
130 34 scopus
135 34 scopus
140 34 scopus
145 34 scopus
150 34 scopus
155 35 scopus
160 36 scopus
165 38 scopus
170 39 scopus
175 39 scopus
180 40 scopus
185 41 scopus
190 41 scopus
195 42 scopus
200 43 scopus
205 44 scopus
210 45 scopus
215 45 scopus
220 46 scopus
225 47 scopus
230 48 scopus
235 48 scopus
240 48 scopus
245 48 scopus
250 50 scopus
255 50 scopus
260 50 scopus
265 51 scopus
270 51 scopus
275 52 scopus
280 52 scopus
285 53 scopus
290 54 scopus
295 55 scopus
300 55 scopus
305 56 scopus
310 56 scopus
315 56 scopus
320 57 scopus
325 57 scopus
330 58 scopus
335 59 scopus
340 59 scopus
345 59 scopus
350 60 scopus
355 61 scopus
360 61 scopus
365 61 scopus
370 61 scopus
375 62 scopus
380 62 scopus
385 62 scopus
390 62 scopus
395 64 scopus
400 64 scopus
405 64 scopus
410 65 scopus
415 66 scopus
420 66 scopus
425 66 scopus
430 67 scopus
435 67 scopus
440 67 scopus
445 67 scopus
450 68 scopus
455 68 scopus
460 70 scopus
465 71 scopus
470 71 scopus
475 73 scopus
480 74 scopus
485 76 scopus
490 77 scopus
495 78 scopus
500 78 scopus
505 79 scopus
510 80 scopus
515 81 scopus
520 82 scopus
525 83 scopus
530 84 scopus
535 84 scopus
540 84 scopus
545 84 scopus
550 84 scopus
555 85 scopus
560 85 scopus
565 86 scopus
570 86 scopus
575 87 scopus
580 87 scopus
585 87 scopus
590 89 scopus
595 89 scopus
600 89 scopus
605 90 scopus
610 90 scopus
615 92 scopus
620 92 scopus
625 92 scopus
630 93 scopus
635 93 scopus
640 93 scopus
645 93 scopus
650 95 scopus
655 95 scopus
660 96 scopus
665 96 scopus
670 96 scopus
675 96 scopus
680 97 scopus
685 97 scopus
690 97 scopus
695 98 scopus
700 98 scopus
705 99 scopus
710 99 scopus
715 100 scopus
720 101 scopus
725 101 scopus
730 102 scopus
735 103 scopus
740 103 scopus
745 104 scopus
750 105 scopus
755 105 scopus
760 105 scopus
765 106 scopus
770 106 scopus
775 106 scopus
780 106 scopus
785 106 scopus
790 106 scopus
795 106 scopus
800 107 scopus
805 107 scopus
810 107 scopus
815 107 scopus
820 107 scopus
825 107 scopus
830 108 scopus
835 108 scopus
840 109 scopus
845 109 scopus
850 109 scopus
855 109 scopus
860 110 scopus
865 110 scopus
870 111 scopus
875 112 scopus
880 113 scopus
885 114 scopus
890 114 scopus
895 115 scopus
900 115 scopus
905 116 scopus
910 117 scopus
915 118 scopus
920 118 scopus
925 118 scopus
930 119 scopus
935 120 scopus
940 120 scopus
945 121 scopus
950 121 scopus
955 121 scopus
960 121 scopus
965 121 scopus
970 122 scopus
975 123 scopus
980 123 scopus
985 123 scopus
990 124 scopus
995 124 scopus
1000 124 scopus
} node[right] {Scopus};
      \addplot+[orange] table[x=x,y=y,meta=l]{
x y l
5 0 google
10 1 google
15 1 google
20 1 google
25 1 google
30 2 google
35 3 google
40 3 google
45 3 google
50 3 google
55 3 google
60 4 google
65 4 google
70 4 google
75 4 google
80 4 google
85 4 google
90 4 google
95 4 google
100 4 google
105 5 google
110 6 google
115 7 google
120 7 google
125 7 google
130 7 google
135 7 google
140 7 google
145 8 google
150 8 google
155 8 google
160 9 google
165 10 google
170 10 google
175 10 google
180 10 google
185 10 google
190 10 google
195 10 google
200 10 google
205 10 google
210 10 google
215 10 google
220 10 google
225 10 google
230 10 google
235 10 google
240 10 google
245 10 google
250 10 google
255 10 google
260 10 google
265 10 google
270 10 google
275 10 google
280 10 google
285 11 google
290 11 google
295 12 google
300 12 google
305 12 google
310 12 google
315 12 google
320 12 google
325 12 google
330 12 google
335 12 google
340 12 google
345 12 google
350 12 google
355 12 google
360 12 google
365 12 google
370 12 google
375 12 google
380 13 google
385 13 google
390 14 google
395 14 google
400 14 google
405 14 google
410 14 google
415 14 google
420 14 google
425 14 google
430 15 google
435 15 google
440 15 google
445 15 google
450 15 google
455 15 google
460 15 google
465 16 google
470 16 google
475 16 google
480 16 google
485 16 google
490 16 google
495 16 google
500 16 google
505 16 google
510 17 google
515 17 google
520 17 google
525 17 google
530 17 google
535 17 google
540 17 google
545 17 google
550 17 google
555 17 google
560 17 google
565 17 google
570 17 google
575 17 google
580 17 google
585 17 google
590 17 google
595 17 google
600 17 google
605 17 google
610 17 google
615 17 google
620 17 google
625 17 google
630 17 google
635 17 google
640 17 google
645 17 google
650 17 google
655 17 google
660 17 google
665 17 google
670 17 google
675 17 google
680 18 google
685 18 google
690 18 google
695 18 google
700 18 google
705 18 google
710 18 google
715 18 google
720 18 google
725 18 google
730 18 google
735 18 google
740 18 google
745 18 google
750 18 google
755 18 google
760 19 google
765 19 google
770 19 google
775 19 google
780 19 google
785 19 google
790 19 google
795 20 google
800 20 google
805 21 google
810 21 google
815 21 google
820 21 google
825 21 google
830 21 google
835 21 google
840 21 google
845 21 google
850 21 google
855 22 google
860 22 google
865 22 google
870 22 google
875 22 google
880 22 google
885 22 google
890 22 google
895 22 google
900 23 google
905 23 google
910 24 google
915 24 google
920 24 google
925 24 google
930 25 google
935 25 google
940 25 google
945 25 google
950 25 google
955 25 google
960 25 google
965 25 google
970 25 google
975 26 google
980 26 google
985 26 google
990 26 google
995 26 google
1000 26 google
} node[right] {Scholar};
      \addplot+[purple] table[x=x,y=y,meta=l]{
x y l
5 0 arxiv
10 0 arxiv
15 0 arxiv
20 0 arxiv
25 0 arxiv
30 0 arxiv
35 0 arxiv
40 0 arxiv
45 0 arxiv
50 0 arxiv
55 0 arxiv
60 0 arxiv
65 0 arxiv
70 0 arxiv
75 0 arxiv
80 0 arxiv
85 0 arxiv
90 0 arxiv
95 0 arxiv
100 0 arxiv
105 0 arxiv
110 0 arxiv
115 0 arxiv
120 0 arxiv
125 0 arxiv
130 0 arxiv
135 0 arxiv
140 0 arxiv
145 0 arxiv
150 0 arxiv
155 0 arxiv
160 0 arxiv
165 0 arxiv
170 0 arxiv
175 1 arxiv
180 1 arxiv
185 1 arxiv
190 1 arxiv
195 1 arxiv
200 1 arxiv
205 1 arxiv
210 1 arxiv
215 1 arxiv
220 1 arxiv
225 1 arxiv
230 1 arxiv
235 1 arxiv
240 1 arxiv
245 1 arxiv
250 1 arxiv
255 1 arxiv
260 1 arxiv
265 1 arxiv
270 1 arxiv
275 1 arxiv
280 1 arxiv
285 1 arxiv
290 1 arxiv
295 1 arxiv
300 1 arxiv
305 1 arxiv
310 1 arxiv
315 1 arxiv
320 1 arxiv
325 1 arxiv
330 1 arxiv
335 1 arxiv
340 1 arxiv
345 1 arxiv
350 1 arxiv
355 1 arxiv
360 1 arxiv
365 1 arxiv
370 1 arxiv
375 1 arxiv
380 1 arxiv
385 1 arxiv
390 1 arxiv
395 1 arxiv
400 1 arxiv
405 2 arxiv
410 2 arxiv
415 2 arxiv
420 2 arxiv
425 2 arxiv
430 2 arxiv
435 2 arxiv
440 2 arxiv
445 2 arxiv
450 2 arxiv
455 2 arxiv
460 2 arxiv
465 2 arxiv
470 2 arxiv
475 2 arxiv
480 2 arxiv
485 2 arxiv
490 2 arxiv
495 2 arxiv
500 2 arxiv
505 2 arxiv
510 2 arxiv
515 2 arxiv
520 2 arxiv
525 3 arxiv
530 3 arxiv
535 3 arxiv
540 3 arxiv
545 4 arxiv
550 4 arxiv
555 4 arxiv
560 4 arxiv
565 4 arxiv
570 4 arxiv
575 4 arxiv
580 4 arxiv
585 4 arxiv
590 4 arxiv
595 4 arxiv
600 4 arxiv
605 4 arxiv
610 4 arxiv
615 4 arxiv
620 4 arxiv
625 4 arxiv
630 4 arxiv
635 4 arxiv
640 4 arxiv
645 4 arxiv
650 4 arxiv
655 4 arxiv
660 4 arxiv
665 4 arxiv
670 4 arxiv
675 4 arxiv
680 5 arxiv
685 5 arxiv
690 5 arxiv
695 5 arxiv
700 5 arxiv
705 5 arxiv
710 5 arxiv
715 5 arxiv
720 5 arxiv
725 5 arxiv
730 6 arxiv
735 6 arxiv
740 6 arxiv
745 6 arxiv
750 6 arxiv
755 6 arxiv
760 6 arxiv
765 6 arxiv
770 6 arxiv
775 6 arxiv
780 6 arxiv
785 7 arxiv
790 8 arxiv
795 8 arxiv
800 8 arxiv
805 8 arxiv
810 8 arxiv
815 8 arxiv
820 8 arxiv
825 8 arxiv
830 8 arxiv
835 8 arxiv
840 8 arxiv
845 8 arxiv
850 8 arxiv
855 8 arxiv
860 9 arxiv
865 10 arxiv
870 10 arxiv
875 10 arxiv
880 10 arxiv
885 10 arxiv
890 10 arxiv
895 10 arxiv
900 10 arxiv
905 10 arxiv
910 10 arxiv
915 10 arxiv
920 10 arxiv
925 10 arxiv
930 10 arxiv
935 10 arxiv
940 10 arxiv
945 10 arxiv
950 10 arxiv
955 10 arxiv
960 10 arxiv
965 10 arxiv
970 10 arxiv
975 10 arxiv
980 10 arxiv
985 10 arxiv
990 10 arxiv
995 10 arxiv
1000 10 arxiv
      } node[right] {ArXiV};
    \end{axis}
  \end{tikzpicture}
  \caption{Cumulative number of unique relevant papers found in each digital library (across all five non-functional property keyword groups).}
  \label{figure:post:lib}

%% file: figure/venn.tex
  \centering
  \begin{tikzpicture}[set/.style={fill=cyan,fill opacity=0.15}]
    \draw[set,
        xshift=-2.7cm,
        yshift=-0.1cm] (0,0) circle (0.6cm);
    \node[align=center] at (-3.3,0.9) {Scholar\\(26)};
    \node at (-2.9,-0.1) {25};

    \draw[set,
        rotate=45] (0,0) ellipse (2cm and 1cm);
    \node[align=center] at (-2.2,1.4) {ACM\\(87)};
    \node at (-1.6,0) {60};
    \node at (-2.3,-0.1) {1};

    \draw[set,
        rotate=-45] (0,0) ellipse (2cm and 1cm);
    \node[align=center] at (-1.1,2) {IEEE\\(105)};
    \node at (-1,1) {75};
    \node at (-1.3,0.5) {1};

    \draw[set,
        yshift=1.4cm] (0,0) circle (0.6cm);
    \node[align=center] at (0,2.5) {ArXiV\\(10)};
    \node at (0,1.6) {9};
    \node at (-0.4,1.2) {0};
    \node at (0,1) {1};

    \draw[set,
        xshift=1cm,
        yshift=-0.705cm,
        rotate=45] (0,0) ellipse (2cm and 1cm);
    \node[align=center] at (1.1,2) {Scopus\\(126)};
    \node at (1,1) {80};
    \node at (0,0.25) {23};
    \node at (0.4,1.2) {0};
    \node at (-0.7,-0.25) {0};
    \node at (-1,-1) {11};

    \draw[set,
        xshift=-1cm,
        yshift=-0.705cm,
        rotate=-45,] (0,0) ellipse (2cm and 1cm);
    \node[align=center] at (2.2,1.4) {Manual\\(100)};
    \node at (2,0) {81};
    \node at (1.3,0.5) {2};
    \node at (0.7,-0.25) {0};
    \node at (0,-0.8) {2};
    \node at (1,-1) {3};
    \node at (-0.35,-1.25) {7};
    \node at (0.35,-1.25) {0};
    \node at (0,-1.8) {5};

  \end{tikzpicture}
  \caption{Venn diagram of all 386 corpus papers according to origin. Missing intersections (e.g., between Google Scholar and ArXiV) are all empty.}
  \label{figure:venn}

%% file: table/papers_summary.tex
\begin{table*}[tp!]
  \caption{%
Summary of the 386 papers of our corpus, according to the main types of fitness function targeted, the types of multi-objective focuses, the types of search approaches, and the types of modifications applied to the original software.
An interactive and more comprehensive artefact for this data is available both as an ancillary file and live at \url{https://bloa.github.io/nfunc_survey}
}
  \centering
  \footnotesize
  \begin{tabular}{p{3em}p{44em}}
    \toprule
    Criteria & Relevant papers\\
    \midrule
    Fitness & \textbf{time~(241)}: \cite{karlin:2012:sc-1,majd:mcsoc:2019,yotov:2003:pldi,liu:2010:ics,mckinley:1998:tpds,voss:2000:icpp,lee:2004:scopes,lu:2009:hpcc,yang:2010:pldi,kayraklioglu:2019:ccgrid,kayraklioglu:2021:tpds,kelefouras:2019:jsc,kelefouras:2018:cf,gerard:2012:lctes,matsurama:2020:corr,selakovic:2017:issta,joshi:2014:icdcs,gupta:2000:date,carvalho:2018:sac,yang:2014:arcs-w,zhou:2022:tpds,hung:2009:icess,jangda:2018:ppopp,alkandari:2021:access,hecht:2016:mobilesoft,chavarria-miranda:2002:pact,psarris:2004:tpds,jingu:2018:candar-w,chen:2019:ase,ma:2010:hipc,ma:2010:hipc,beach:2009:uchpc-maw,arabnejad:2018:bdcloud,wilcox:2014:jsep,sandran:2011:socpros,desai:2009:iadcc,zhou:2011:pldi,arcelli:2012:qosa,hernandez:2008:sp,langdon:2016:ssbse,gibson:2000:icsm,lopez-lopez:2019:softc,petke:2013:ssbse,cody-kenny:2017:gecco,heydemann:2005:scopes,jia:2013:qsic:1,deng:2021:iccrd,zhu:2001:ics,joisha:2011:popl,brumar:2017:ipdps,ren:2008:pact,kandemir:2000:tpds-1,aleen:2016:cf,rahman:2011:hipeac,hoste:2010:cgo,moesus:2018:icsoft,malecha:2015:sac,czako:2021:esa,dave:2009:lcpc,djoudi:2008:saw,milosavljevic:1999:ipps-spdp,jang:2012:taco,werner:2020:mlcad,yuki:2010:cgo,baskaran:2010:cc,jin:2000:ishpc,spazier:2016:array-pldi,kocsis:2016:gi-gecco,danalis:2007:ipdps,kalms:2018:parma-ditam,seo:2013:pact,lee:2011:cgo,varrette:2019:ppam,noel:1998:iccl,kumar:2021:corr,mesmay:2009:icml,tilevich:2005:icse,desani:2016:pdp,pachenko:2019:cgo,satish:2012:isca,farzat:2018:jserd,ashouri:2016:taco,cho:2009:asp-dac,wang:2007:cgo:1,ramirez:2001:isca,williams-king:2019:feast-ccs,shi:2013:oopsla,lee:2010:interact-asplos,ploensin:2021:icapm,pai:1999:micro,bhattacharya:2013:oopsla,rele:2001:tcad,blot:2020:eurogp,cooper:1998:asplos,azeemi:2006:icet,yemliha:2007:vlsid,son:2007:tpds,kandemir:2001:codes,nezzari:2017:aero,zhai:2004:cgo,triantafyllis:2003:cgo,barua:2001:tc,ishizaki:2015:pact,lima:2013:patmos,bolat:2009:ipccc,li:2019:asap,basios:2018:fse,koc:2020:ccwc,lakhdar:2020:scopes,wu:2015:gecco,weidmann:1997:iscope,bagge:2003:scam,ravishankar:2015:ppopp,leopoldseder:2018:cgo,che:2011:sc,ghanim:2018:tpds,tariq:2020:icacs,coplin:2015:gpgpu-ppopp,rasch:2021:taco,chen:2021:icse,sharma:2000:vlsid,blot:2021:tevc,rawat:2015:date,guizzo:2021:icse,badawi:2001:ics,white:2011:tevc,garciarena:2016:gi-gecco,langdon:2010:cec,sato:2015:seps-splash:1,sarkar:2014:compute,moreton-fernandez:2014:hpcs,koch:2014:lctes,magni:2014:gpgpu-asplos,lin:2019:corr,welton:2018:ccgrid,maras:2012:www,pan:2006:cgo,tagliavini:2020:tcad,leupers:1999:iccad,dathathri:2013:pact,liou:2019:gi-icse,langdon:2017:gpem,haraldsson:2017:gi-gecco:2,liou:2020:acm-taco,gu:2020:hipc,fang:2014:icpp,langdon:2015:gi-gecco,feld:2015:cosmic-cgo,ma:2016:jcst,holewinski:2012:ics,jackson:2011:pdp:1,hagedorn:2018:cgo,sahin:2014:esem,welton:2020:ics,branco:2015:isci,sedgewick:1978:cacm,heid:2018:fsp-fpl,langdon:2014:gecco,langdon:2015:gecco,langdon:2017:gi-gecco,carr:1994:tpls,brownlee:2020:cec,neth:2021:ics,barik:2013:pact,papadopoulos:2018:scopes,sachan:2021:jsa,alcocer:2016:icpe,cody-kenny:2015:gi-gecco,martonosi:1992:sigmetrics,tsafrir:2009:oopsla,colucci:2020:corr,blot:2016:lion,azeemi:2006:iit,umeda:2015:lcpc,lopez-lopez:2018:gi-gecco,wang:2010:cgo,haneda:2006:ipdps,kumar:2003:icvd,ren:2010:dsde,canales:2021:icpe-wip,fischer:2006:hpcmp-ugc,haber:2003:cgo,ryoo:2008:ppopp,langdon:2015:tevc,cooper:1999:lctes,akiba:2019:kdd,doeraene:2016:oopsla,asenjo:2008:ipdps,hutter:2009:jair,pinto:2022:tse,cody-kenny:2018:gi-icse,sandran:2012:cec,bondhugula:2008:pldi,schulte:2014:asplos,karlin:2012:sc-2,hennessy:1983:tpls,ball:1979:cc,debray:2002:pldi,amidon:2015:hpec,sanguinetti:1984:sigmetrics,leon:2016:parco,an:2019:fse,kadayif:2004:tpds,cavazos:2007:cgo,che:2005:hpcasia,su:2019:icse,lyu:2018:issta,brandvein:2016:pepm,hung-cuong:2013:iccasa,devuyst:2011:hipeac,khatchadourian:2019:icse,xu:2014:acm-tosem,girbal:2006:ijpp,winterstein:2014:fccm,hutter:2011:lion,scales:1996:asplos,vallee-rai:2010:cascon-fdhip,darulova:2018:iccps,simunic:2000:isss,chung:2001:isss,zhou:2014:oopsla,petke:2018:tse,adler:2014:spe,lai:2015:emsoft,rahman:2012:cf,stratis:2016:ase,chan:1994:can,ierotheou:2001:sp,abdelaal:2021:ics,rivera:2000:sc,linderman:2010:cgo,wu:2020:ics,brunie:2020:sc,weng:2021:corr,mcphee:2017:gecco-c_1,ierotheou:2007:ijcm,petke:2014:eurogp,park:2002:asia-pepm,chung:2002:tcad,kanada:1988:ics,kulkarni:2006:tecs}, \textbf{energy~(71)}: \cite{delaluz:2004:tpds,zou:2018:tpds,roy:2007:vlsid,kelefouras:2019:jsc,luo:2009:emc,ukidave:2013:ashes-ipdps,bruce:2019:tse,shih:2016:icpp-w,chandar:2006:jsps,chandar:2006:jsps,li:2016:scam,fernandes:2017:icse-c,rahman:2011:hipeac,dorn:2019:tse,varrette:2019:ppam,cong:2009:islped,kim:2018:cc,kulkarni:1998:ipps-spdp,ibrahim:2009:newcas,azeemi:2006:icet,son:2007:tpds,lima:2013:patmos,cruz:2019:ecsme,bhattacharya:2012:sigmetrics,guzma:2011:soc,coplin:2015:gpgpu-ppopp,chen:2016:jpdc,fei:2007:tecs,fei:2004:icvd,anwar:2019:seaa,jakobs:2018:hpcs,roy:2009:iscas,sehgal:2020:jksu-cis,sahin:2014:esem,branco:2015:isci,nguyen:2016:iukm,bree:2020:iwor-icse,carette:2017:saner,park:2014:seke,bokhari:2018:mobiquitous,pereira:2018:ase,sachan:2021:jsa,li:2014:icse,nikzad:2015:middleware,colucci:2020:corr,azeemi:2006:iit,barati:2021:corr_1,kambadur:2016:cgo,burles:2015:ssbse-1,li:2015:demobile-fse:1,pinto:2022:tse,cruz:2017:mobilesoft:1,schulte:2014:asplos,honig:2014:trios,abdusalam:2014:igcc,leon:2016:parco,oliveira:2019:msr,bruce:2015:gecco,lyu:2018:issta,brownlee:2017:tetci,gutierrez:2014:icse,tan:2003:date,simunic:2000:isss,chung:2001:isss,canino:2018:fse,rahman:2012:cf,sakamoto:2020:hpcasia,daud:2009:ijict,pereira:2016:greens-icse,cruz:2017:mobilesoft:2,chung:2002:tcad}, \textbf{size~(69)}: \cite{seong:2006:iccad,soto-valero:2021:ese,lee:2004:scopes,marino:2016:ppsn,courbot:2006:cardis,halambi:2002:date,chabbi:2021:cgo,chandar:2006:jsps,chandar:2006:jsps,heydemann:2005:scopes,jia:2013:qsic:1,bosch:2014:doceng,malecha:2015:sac,werner:2020:mlcad,mururu:2019:corr,porter:2020:pldi,quach:2019:feast-ccs,brown:2019:feast-ccs,rastogi:2017:fse,avgerinos:2009:erlang,brown:2010:pepm,cho:2009:asp-dac,drinick:2003:cgo,lee:2010:interact-asplos,zhuge:2003:tecs,biswas:2021:codaspy,sutter:2001:om-pldi,chebolu:2016:icacci,geetha:2008:adcom,koo:2019:eurosec,rajan:1999:codes,ghanim:2018:tpds,heo:2018:css,yeboah-antwi:2015:gi-gecco,breternitz:1997:iccd,thangaraj:2006:adcom,farzat:2021:tse,koch:2014:lctes,badri:2017:icmla,dasilva:2021:cc,maras:2012:www,bonny:2008:iccad,leupers:1999:iccad,heikkinen:2003:issoc,lefurgy:1997:micro,med:2007:icsamos,jiang:2016:compsac,bruce:2020:fse,yeboah-antwi:2017:gpem,xiao:2002:isss,zhuge:2002:icpp,schulte:2014:asplos,hennessy:1983:tpls,debray:2002:pldi,xin:2020:icse-nier,qian:2019:uss,stefanov:2004:ifl,qian:2020:ccs,vasquez:2019:ist,jiang:2022:tr,lai:2015:emsoft,xin:2020:ase,yang:2012:trustcom,plumbridge:2012:recosoc,ahmad:2022:tse,sharif:2018:ase,xing:2022:tmc,chung:2002:tcad,macias:2020:fse}, \textbf{other~(54)}: \cite{shi:2015:cal,nasagh:2021:softc,wilcox:2014:jsep,eldib:2013:fmcad,lopez-lopez:2019:softc,jeon:2012:racs,brumar:2017:ipdps,dorn:2019:tse,ghaffarinia:2019:ccs,mururu:2019:corr,porter:2020:pldi,bergel:2005:oopsla,nezzari:2017:aero,koo:2019:eurosec,fatima:2020:quasoq-apsec,cruz:2019:ecsme,lucas:2019:sbes,heo:2018:css,cherian:2020:corr,chionis:2013:pci,white:2011:tevc,blaha:2002:cec,langdon:2020:gi-gecco,langdon:2019:gecco-comp-2,langdon:2019:gi-gecco,liou:2019:gi-icse,lopez-lopez:2016:gi-gecco,liou:2020:acm-taco,papadopoulos:2018:scopes,bokhari:2018:mobiquitous,brown:2019:cset,wang:2019:access-2,blot:2016:lion,devore-mcdonald:2020:oopsla,barati:2021:corr_1,agadakos:2019:acsac,lopez-lopez:2018:gi-gecco,linn:2003:ccs,bartoli:2014:gecco,xin:2020:icse-nier,tao:2014:icsme,duan:2020:secdev,larus:1988:ppeals,khatchadourian:2019:icse,brownlee:2017:tetci,berthelon:2013:nidisc-ipdps,guo:2020:corr,xin:2020:ase,ghavamnia:2020:uss,haraldsson:2017:iscc,poor:2020:erlang-icfp,sharif:2018:ase,moyasebi:2020:gi-gecco,mcphee:2017:gecco-c_1}, \textbf{memory~(35)}: \cite{strobel:2019:fdl,mathur:2015:candar,hwu:1989:isca:2,kandemir:2004:tcad,kelefouras:2019:jsc,gerard:2012:lctes,joshi:2014:icdcs,alkandari:2021:access,hecht:2016:mobilesoft,chavarria-miranda:2002:pact,delaluz:2004:tc,jia:2013:qsic:1,yan:2016:sac,ghosh:1999:tpls,kulkarni:1998:ipps-spdp,yemliha:2007:vlsid,kandemir:2001:codes,barua:2001:tc,basios:2018:fse,wu:2015:gecco,bagge:2003:scam,bhattacharya:2012:sigmetrics,ilham:2011:iceei,hamza:2013:pppj,viessman:2018:ifl,grunwald:1993:pldi,papadopoulos:2018:scopes,katsaragakis:2020:iscas,martonosi:1992:sigmetrics,colucci:2020:corr,blot:2016:lion,karlin:2012:sc-2,bhattacharya:2011:ecoop,xu:2014:acm-tosem,lai:2015:emsoft} \\
    Multi-objective & \textbf{focus~(57)}: \cite{lee:2004:scopes,kelefouras:2019:jsc,gerard:2012:lctes,joshi:2014:icdcs,alkandari:2021:access,hecht:2016:mobilesoft,chavarria-miranda:2002:pact,lopez-lopez:2019:softc,chandar:2006:jsps,chandar:2006:jsps,jia:2013:qsic:1,brumar:2017:ipdps,malecha:2015:sac,mururu:2019:corr,porter:2020:pldi,cho:2009:asp-dac,lee:2010:interact-asplos,kulkarni:1998:ipps-spdp,yemliha:2007:vlsid,son:2007:tpds,kandemir:2001:codes,nezzari:2017:aero,barua:2001:tc,koo:2019:eurosec,bagge:2003:scam,cruz:2019:ecsme,bhattacharya:2012:sigmetrics,ghanim:2018:tpds,heo:2018:css,coplin:2015:gpgpu-ppopp,koch:2014:lctes,maras:2012:www,leupers:1999:iccad,sahin:2014:esem,branco:2015:isci,papadopoulos:2018:scopes,sachan:2021:jsa,martonosi:1992:sigmetrics,colucci:2020:corr,schulte:2014:asplos,karlin:2012:sc-2,hennessy:1983:tpls,debray:2002:pldi,xin:2020:icse-nier,leon:2016:parco,lyu:2018:issta,khatchadourian:2019:icse,xu:2014:acm-tosem,simunic:2000:isss,chung:2001:isss,lai:2015:emsoft,rahman:2012:cf,xin:2020:ase,poor:2020:erlang-icfp,sharif:2018:ase,mcphee:2017:gecco-c_1,chung:2002:tcad}, \textbf{report~(54)}: \cite{strobel:2019:fdl,mathur:2015:candar,delaluz:2004:tpds,shi:2015:cal,kayraklioglu:2019:ccgrid,kayraklioglu:2021:tpds,luo:2009:emc,ukidave:2013:ashes-ipdps,delaluz:2004:tc,zhu:2001:ics,joisha:2011:popl,lee:2011:cgo,kumar:2021:corr,ghaffarinia:2019:ccs,brown:2019:feast-ccs,shi:2013:oopsla,ibrahim:2009:newcas,cooper:1998:asplos,leopoldseder:2018:cgo,guzma:2011:soc,sharma:2000:vlsid,chionis:2013:pci,ilham:2011:iceei,farzat:2021:tse,jakobs:2018:hpcs,hamza:2013:pppj,badri:2017:icmla,viessman:2018:ifl,tagliavini:2020:tcad,feld:2015:cosmic-cgo,grunwald:1993:pldi,jiang:2016:compsac,pereira:2018:ase,wang:2019:access-2,nikzad:2015:middleware,linn:2003:ccs,haber:2003:cgo,langdon:2015:tevc,ball:1979:cc,honig:2014:trios,kadayif:2004:tpds,su:2019:icse,duan:2020:secdev,brandvein:2016:pepm,hung-cuong:2013:iccasa,girbal:2006:ijpp,scales:1996:asplos,qian:2020:ccs,jiang:2022:tr,adler:2014:spe,daud:2009:ijict,rivera:2000:sc,plumbridge:2012:recosoc,ahmad:2022:tse}, \textbf{search~(22)}: \cite{wilcox:2014:jsep,heydemann:2005:scopes,rahman:2011:hipeac,hoste:2010:cgo,dorn:2019:tse,werner:2020:mlcad,varrette:2019:ppam,azeemi:2006:icet,lima:2013:patmos,basios:2018:fse,wu:2015:gecco,white:2011:tevc,liou:2019:gi-icse,lopez-lopez:2016:gi-gecco,liou:2020:acm-taco,bokhari:2018:mobiquitous,blot:2016:lion,azeemi:2006:iit,barati:2021:corr_1,lopez-lopez:2018:gi-gecco,pinto:2022:tse,brownlee:2017:tetci} \\
    Search & \textbf{static~(204)}: \cite{strobel:2019:fdl,seong:2006:iccad,delaluz:2004:tpds,zou:2018:tpds,liu:2010:ics,roy:2007:vlsid,mckinley:1998:tpds,soto-valero:2021:ese,shi:2015:cal,yang:2010:pldi,courbot:2006:cardis,kelefouras:2019:jsc,kelefouras:2018:cf,gerard:2012:lctes,matsurama:2020:corr,selakovic:2017:issta,joshi:2014:icdcs,luo:2009:emc,gupta:2000:date,ukidave:2013:ashes-ipdps,zhou:2022:tpds,jangda:2018:ppopp,halambi:2002:date,alkandari:2021:access,chavarria-miranda:2002:pact,chabbi:2021:cgo,psarris:2004:tpds,jingu:2018:candar-w,ma:2010:hipc,ma:2010:hipc,arabnejad:2018:bdcloud,eldib:2013:fmcad,zhou:2011:pldi,hernandez:2008:sp,gibson:2000:icsm,shih:2016:icpp-w,chandar:2006:jsps,chandar:2006:jsps,jeon:2012:racs,delaluz:2004:tc,li:2016:scam,fernandes:2017:icse-c,deng:2021:iccrd,zhu:2001:ics,joisha:2011:popl,brumar:2017:ipdps,kandemir:2000:tpds-1,aleen:2016:cf,yan:2016:sac,bosch:2014:doceng,moesus:2018:icsoft,malecha:2015:sac,milosavljevic:1999:ipps-spdp,jang:2012:taco,baskaran:2010:cc,jin:2000:ishpc,spazier:2016:array-pldi,kocsis:2016:gi-gecco,danalis:2007:ipdps,kalms:2018:parma-ditam,lee:2011:cgo,noel:1998:iccl,kumar:2021:corr,cong:2009:islped,ghaffarinia:2019:ccs,mururu:2019:corr,tilevich:2005:icse,porter:2020:pldi,quach:2019:feast-ccs,pachenko:2019:cgo,ghosh:1999:tpls,satish:2012:isca,brown:2019:feast-ccs,rastogi:2017:fse,avgerinos:2009:erlang,cho:2009:asp-dac,wang:2007:cgo:1,ramirez:2001:isca,williams-king:2019:feast-ccs,drinick:2003:cgo,shi:2013:oopsla,lee:2010:interact-asplos,zhuge:2003:tecs,ploensin:2021:icapm,kulkarni:1998:ipps-spdp,pai:1999:micro,bhattacharya:2013:oopsla,sutter:2001:om-pldi,rele:2001:tcad,cooper:1998:asplos,yemliha:2007:vlsid,son:2007:tpds,kandemir:2001:codes,nezzari:2017:aero,zhai:2004:cgo,barua:2001:tc,ishizaki:2015:pact,geetha:2008:adcom,koo:2019:eurosec,li:2019:asap,koc:2020:ccwc,lakhdar:2020:scopes,bagge:2003:scam,fatima:2020:quasoq-apsec,rajan:1999:codes,ravishankar:2015:ppopp,leopoldseder:2018:cgo,che:2011:sc,ghanim:2018:tpds,tariq:2020:icacs,sharma:2000:vlsid,rawat:2015:date,breternitz:1997:iccd,thangaraj:2006:adcom,badawi:2001:ics,chionis:2013:pci,jakobs:2018:hpcs,sato:2015:seps-splash:1,sarkar:2014:compute,moreton-fernandez:2014:hpcs,koch:2014:lctes,roy:2009:iscas,lin:2019:corr,viessman:2018:ifl,maras:2012:www,bonny:2008:iccad,dathathri:2013:pact,fang:2014:icpp,feld:2015:cosmic-cgo,hagedorn:2018:cgo,welton:2020:ics,heikkinen:2003:issoc,heid:2018:fsp-fpl,nguyen:2016:iukm,lefurgy:1997:micro,neth:2021:ics,barik:2013:pact,carette:2017:saner,park:2014:seke,brown:2019:cset,jiang:2016:compsac,bruce:2020:fse,pereira:2018:ase,wang:2019:access-2,nikzad:2015:middleware,katsaragakis:2020:iscas,umeda:2015:lcpc,agadakos:2019:acsac,linn:2003:ccs,wang:2010:cgo,kumar:2003:icvd,xiao:2002:isss,zhuge:2002:icpp,ren:2010:dsde,haber:2003:cgo,ryoo:2008:ppopp,li:2015:demobile-fse:1,doeraene:2016:oopsla,asenjo:2008:ipdps,bondhugula:2008:pldi,hennessy:1983:tpls,honig:2014:trios,debray:2002:pldi,amidon:2015:hpec,kadayif:2004:tpds,qian:2019:uss,oliveira:2019:msr,che:2005:hpcasia,su:2019:icse,tao:2014:icsme,lyu:2018:issta,brandvein:2016:pepm,larus:1988:ppeals,bhattacharya:2011:ecoop,hung-cuong:2013:iccasa,devuyst:2011:hipeac,xu:2014:acm-tosem,girbal:2006:ijpp,winterstein:2014:fccm,scales:1996:asplos,stefanov:2004:ifl,qian:2020:ccs,vasquez:2019:ist,guo:2020:corr,vallee-rai:2010:cascon-fdhip,adler:2014:spe,lai:2015:emsoft,ghavamnia:2020:uss,yang:2012:trustcom,abdelaal:2021:ics,rivera:2000:sc,linderman:2010:cgo,poor:2020:erlang-icfp,plumbridge:2012:recosoc,ahmad:2022:tse,sharif:2018:ase,xing:2022:tmc,wu:2020:ics,weng:2021:corr,ierotheou:2007:ijcm,cruz:2017:mobilesoft:2,park:2002:asia-pepm,kanada:1988:ics,macias:2020:fse}, \textbf{evolutionary~(75)}: \cite{majd:mcsoc:2019,lu:2009:hpcc,nasagh:2021:softc,marino:2016:ppsn,kayraklioglu:2019:ccgrid,kayraklioglu:2021:tpds,hung:2009:icess,beach:2009:uchpc-maw,sandran:2011:socpros,langdon:2016:ssbse,lopez-lopez:2019:softc,petke:2013:ssbse,bruce:2019:tse,cody-kenny:2017:gecco,hoste:2010:cgo,czako:2021:esa,dorn:2019:tse,varrette:2019:ppam,farzat:2018:jserd,chebolu:2016:icacci,blot:2020:eurogp,azeemi:2006:icet,basios:2018:fse,wu:2015:gecco,yeboah-antwi:2015:gi-gecco,blot:2021:tevc,chen:2016:jpdc,guizzo:2021:icse,white:2011:tevc,garciarena:2016:gi-gecco,blaha:2002:cec,langdon:2010:cec,farzat:2021:tse,langdon:2020:gi-gecco,dasilva:2021:cc,langdon:2019:gecco-comp-2,langdon:2019:gi-gecco,liou:2019:gi-icse,langdon:2017:gpem,haraldsson:2017:gi-gecco:2,lopez-lopez:2016:gi-gecco,liou:2020:acm-taco,langdon:2015:gi-gecco,langdon:2014:gecco,langdon:2015:gecco,langdon:2017:gi-gecco,bokhari:2018:mobiquitous,sachan:2021:jsa,cody-kenny:2015:gi-gecco,azeemi:2006:iit,barati:2021:corr_1,lopez-lopez:2018:gi-gecco,burles:2015:ssbse-1,yeboah-antwi:2017:gpem,canales:2021:icpe-wip,langdon:2015:tevc,cooper:1999:lctes,akiba:2019:kdd,cody-kenny:2018:gi-icse,sandran:2012:cec,bartoli:2014:gecco,schulte:2014:asplos,cavazos:2007:cgo,bruce:2015:gecco,khatchadourian:2019:icse,brownlee:2017:tetci,hutter:2011:lion,berthelon:2013:nidisc-ipdps,jiang:2022:tr,darulova:2018:iccps,petke:2018:tse,haraldsson:2017:iscc,mcphee:2017:gecco-c_1,petke:2014:eurogp,kulkarni:2006:tecs}, \textbf{manual~(46)}: \cite{karlin:2012:sc-1,mathur:2015:candar,hecht:2016:mobilesoft,chen:2019:ase,bergel:2005:oopsla,brown:2010:pepm,kim:2018:cc,ibrahim:2009:newcas,weidmann:1997:iscope,cruz:2019:ecsme,bhattacharya:2012:sigmetrics,lucas:2019:sbes,anwar:2019:seaa,ilham:2011:iceei,jakobs:2018:hpcs,hamza:2013:pppj,badri:2017:icmla,welton:2018:ccgrid,sehgal:2020:jksu-cis,jackson:2011:pdp:1,hagedorn:2018:cgo,sahin:2014:esem,welton:2020:ics,sedgewick:1978:cacm,carr:1994:tpls,bree:2020:iwor-icse,papadopoulos:2018:scopes,alcocer:2016:icpe,martonosi:1992:sigmetrics,tsafrir:2009:oopsla,kambadur:2016:cgo,fischer:2006:hpcmp-ugc,cruz:2017:mobilesoft:1,karlin:2012:sc-2,ball:1979:cc,abdusalam:2014:igcc,sanguinetti:1984:sigmetrics,leon:2016:parco,duan:2020:secdev,tan:2003:date,simunic:2000:isss,rahman:2012:cf,sakamoto:2020:hpcasia,pereira:2016:greens-icse,ierotheou:2001:sp,ierotheou:2007:ijcm}, \textbf{exploratory~(44)}: \cite{desai:2009:iadcc,ren:2008:pact,rahman:2011:hipeac,dave:2009:lcpc,werner:2020:mlcad,yuki:2010:cgo,mesmay:2009:icml,desani:2016:pdp,ashouri:2016:taco,biswas:2021:codaspy,triantafyllis:2003:cgo,lima:2013:patmos,bolat:2009:ipccc,heo:2018:css,rasch:2021:taco,chen:2021:icse,cherian:2020:corr,blot:2021:tevc,fei:2007:tecs,fei:2004:icvd,farzat:2021:tse,pan:2006:cgo,tagliavini:2020:tcad,leupers:1999:iccad,gu:2020:hipc,brownlee:2020:cec,med:2007:icsamos,li:2014:icse,colucci:2020:corr,blot:2016:lion,devore-mcdonald:2020:oopsla,hutter:2009:jair,pinto:2022:tse,sandran:2012:cec,xin:2020:icse-nier,an:2019:fse,cavazos:2007:cgo,gutierrez:2014:icse,jiang:2022:tr,zhou:2014:oopsla,canino:2018:fse,xin:2020:ase,brunie:2020:sc,chung:2002:tcad}, \textbf{sampling~(27)}: \cite{hwu:1989:isca:2,yotov:2003:pldi,kandemir:2004:tcad,voss:2000:icpp,lee:2004:scopes,carvalho:2018:sac,yang:2014:arcs-w,wilcox:2014:jsep,arcelli:2012:qosa,heydemann:2005:scopes,jia:2013:qsic:1,djoudi:2008:saw,seo:2013:pact,guzma:2011:soc,coplin:2015:gpgpu-ppopp,magni:2014:gpgpu-asplos,ma:2016:jcst,holewinski:2012:ics,branco:2015:isci,grunwald:1993:pldi,haneda:2006:ipdps,pinto:2022:tse,chung:2001:isss,stratis:2016:ase,chan:1994:can,daud:2009:ijict,moyasebi:2020:gi-gecco} \\
    Change & \textbf{semantic~(206)}: \cite{strobel:2019:fdl,seong:2006:iccad,karlin:2012:sc-1,mathur:2015:candar,hwu:1989:isca:2,liu:2010:ics,roy:2007:vlsid,mckinley:1998:tpds,soto-valero:2021:ese,lee:2004:scopes,nasagh:2021:softc,courbot:2006:cardis,gerard:2012:lctes,matsurama:2020:corr,selakovic:2017:issta,luo:2009:emc,zhou:2022:tpds,halambi:2002:date,alkandari:2021:access,hecht:2016:mobilesoft,chabbi:2021:cgo,chen:2019:ase,ma:2010:hipc,ma:2010:hipc,beach:2009:uchpc-maw,arabnejad:2018:bdcloud,wilcox:2014:jsep,eldib:2013:fmcad,zhou:2011:pldi,arcelli:2012:qosa,gibson:2000:icsm,shih:2016:icpp-w,chandar:2006:jsps,chandar:2006:jsps,jeon:2012:racs,heydemann:2005:scopes,li:2016:scam,fernandes:2017:icse-c,deng:2021:iccrd,zhu:2001:ics,joisha:2011:popl,brumar:2017:ipdps,aleen:2016:cf,bosch:2014:doceng,moesus:2018:icsoft,malecha:2015:sac,milosavljevic:1999:ipps-spdp,jang:2012:taco,jin:2000:ishpc,spazier:2016:array-pldi,kocsis:2016:gi-gecco,danalis:2007:ipdps,lee:2011:cgo,noel:1998:iccl,kumar:2021:corr,cong:2009:islped,ghaffarinia:2019:ccs,mururu:2019:corr,tilevich:2005:icse,porter:2020:pldi,quach:2019:feast-ccs,pachenko:2019:cgo,satish:2012:isca,brown:2019:feast-ccs,rastogi:2017:fse,bergel:2005:oopsla,avgerinos:2009:erlang,brown:2010:pepm,cho:2009:asp-dac,wang:2007:cgo:1,ramirez:2001:isca,williams-king:2019:feast-ccs,drinick:2003:cgo,shi:2013:oopsla,kim:2018:cc,lee:2010:interact-asplos,biswas:2021:codaspy,kulkarni:1998:ipps-spdp,sutter:2001:om-pldi,rele:2001:tcad,cooper:1998:asplos,nezzari:2017:aero,zhai:2004:cgo,ishizaki:2015:pact,geetha:2008:adcom,bolat:2009:ipccc,koc:2020:ccwc,lakhdar:2020:scopes,weidmann:1997:iscope,bagge:2003:scam,fatima:2020:quasoq-apsec,rajan:1999:codes,lucas:2019:sbes,leopoldseder:2018:cgo,che:2011:sc,ghanim:2018:tpds,heo:2018:css,tariq:2020:icacs,sharma:2000:vlsid,rawat:2015:date,fei:2007:tecs,fei:2004:icvd,breternitz:1997:iccd,thangaraj:2006:adcom,anwar:2019:seaa,badawi:2001:ics,chionis:2013:pci,jakobs:2018:hpcs,koch:2014:lctes,hamza:2013:pppj,roy:2009:iscas,lin:2019:corr,badri:2017:icmla,welton:2018:ccgrid,viessman:2018:ifl,bonny:2008:iccad,leupers:1999:iccad,sehgal:2020:jksu-cis,fang:2014:icpp,ma:2016:jcst,holewinski:2012:ics,jackson:2011:pdp:1,hagedorn:2018:cgo,sahin:2014:esem,heikkinen:2003:issoc,sedgewick:1978:cacm,nguyen:2016:iukm,lefurgy:1997:micro,bree:2020:iwor-icse,med:2007:icsamos,barik:2013:pact,papadopoulos:2018:scopes,carette:2017:saner,park:2014:seke,brown:2019:cset,jiang:2016:compsac,bruce:2020:fse,pereira:2018:ase,alcocer:2016:icpe,wang:2019:access-2,nikzad:2015:middleware,katsaragakis:2020:iscas,martonosi:1992:sigmetrics,tsafrir:2009:oopsla,umeda:2015:lcpc,barati:2021:corr_1,agadakos:2019:acsac,kambadur:2016:cgo,linn:2003:ccs,kumar:2003:icvd,ren:2010:dsde,fischer:2006:hpcmp-ugc,haber:2003:cgo,li:2015:demobile-fse:1,doeraene:2016:oopsla,asenjo:2008:ipdps,cruz:2017:mobilesoft:1,karlin:2012:sc-2,hennessy:1983:tpls,ball:1979:cc,honig:2014:trios,debray:2002:pldi,xin:2020:icse-nier,abdusalam:2014:igcc,sanguinetti:1984:sigmetrics,leon:2016:parco,qian:2019:uss,oliveira:2019:msr,su:2019:icse,tao:2014:icsme,duan:2020:secdev,lyu:2018:issta,brandvein:2016:pepm,larus:1988:ppeals,bhattacharya:2011:ecoop,devuyst:2011:hipeac,xu:2014:acm-tosem,brownlee:2017:tetci,winterstein:2014:fccm,berthelon:2013:nidisc-ipdps,scales:1996:asplos,stefanov:2004:ifl,qian:2020:ccs,vasquez:2019:ist,tan:2003:date,vallee-rai:2010:cascon-fdhip,darulova:2018:iccps,simunic:2000:isss,zhou:2014:oopsla,adler:2014:spe,lai:2015:emsoft,xin:2020:ase,ghavamnia:2020:uss,yang:2012:trustcom,ierotheou:2001:sp,poor:2020:erlang-icfp,plumbridge:2012:recosoc,ahmad:2022:tse,sharif:2018:ase,xing:2022:tmc,ierotheou:2007:ijcm,cruz:2017:mobilesoft:2,park:2002:asia-pepm,chung:2002:tcad,kanada:1988:ics,macias:2020:fse}, \textbf{loops~(96)}: \cite{karlin:2012:sc-1,delaluz:2004:tpds,majd:mcsoc:2019,zou:2018:tpds,yotov:2003:pldi,kandemir:2004:tcad,shi:2015:cal,voss:2000:icpp,lu:2009:hpcc,yang:2010:pldi,kayraklioglu:2019:ccgrid,kayraklioglu:2021:tpds,kelefouras:2019:jsc,kelefouras:2018:cf,matsurama:2020:corr,joshi:2014:icdcs,luo:2009:emc,gupta:2000:date,ukidave:2013:ashes-ipdps,carvalho:2018:sac,yang:2014:arcs-w,zhou:2022:tpds,jangda:2018:ppopp,chavarria-miranda:2002:pact,psarris:2004:tpds,jingu:2018:candar-w,ma:2010:hipc,ma:2010:hipc,arabnejad:2018:bdcloud,delaluz:2004:tc,ren:2008:pact,kandemir:2000:tpds-1,aleen:2016:cf,rahman:2011:hipeac,dave:2009:lcpc,djoudi:2008:saw,yuki:2010:cgo,baskaran:2010:cc,kalms:2018:parma-ditam,seo:2013:pact,ghosh:1999:tpls,zhuge:2003:tecs,ploensin:2021:icapm,kulkarni:1998:ipps-spdp,pai:1999:micro,rele:2001:tcad,azeemi:2006:icet,yemliha:2007:vlsid,son:2007:tpds,kandemir:2001:codes,barua:2001:tc,ishizaki:2015:pact,li:2019:asap,bagge:2003:scam,rajan:1999:codes,ravishankar:2015:ppopp,guzma:2011:soc,sato:2015:seps-splash:1,sarkar:2014:compute,moreton-fernandez:2014:hpcs,magni:2014:gpgpu-asplos,dathathri:2013:pact,feld:2015:cosmic-cgo,ma:2016:jcst,holewinski:2012:ics,hagedorn:2018:cgo,sedgewick:1978:cacm,heid:2018:fsp-fpl,carr:1994:tpls,neth:2021:ics,kambadur:2016:cgo,xiao:2002:isss,zhuge:2002:icpp,ren:2010:dsde,ryoo:2008:ppopp,pinto:2022:tse,bondhugula:2008:pldi,karlin:2012:sc-2,leon:2016:parco,kadayif:2004:tpds,devuyst:2011:hipeac,khatchadourian:2019:icse,girbal:2006:ijpp,winterstein:2014:fccm,berthelon:2013:nidisc-ipdps,guo:2020:corr,simunic:2000:isss,chung:2001:isss,lai:2015:emsoft,abdelaal:2021:ics,rivera:2000:sc,ahmad:2022:tse,sharif:2018:ase,wu:2020:ics,weng:2021:corr,chung:2002:tcad}, \textbf{destructive~(78)}: \cite{karlin:2012:sc-1,marino:2016:ppsn,ukidave:2013:ashes-ipdps,langdon:2016:ssbse,lopez-lopez:2019:softc,petke:2013:ssbse,bruce:2019:tse,cody-kenny:2017:gecco,yan:2016:sac,dorn:2019:tse,farzat:2018:jserd,shi:2013:oopsla,bhattacharya:2013:oopsla,blot:2020:eurogp,li:2019:asap,basios:2018:fse,wu:2015:gecco,cruz:2019:ecsme,bhattacharya:2012:sigmetrics,coplin:2015:gpgpu-ppopp,yeboah-antwi:2015:gi-gecco,blot:2021:tevc,chen:2016:jpdc,guizzo:2021:icse,white:2011:tevc,blaha:2002:cec,langdon:2010:cec,farzat:2021:tse,langdon:2020:gi-gecco,maras:2012:www,tagliavini:2020:tcad,langdon:2019:gecco-comp-2,langdon:2019:gi-gecco,liou:2019:gi-icse,langdon:2017:gpem,haraldsson:2017:gi-gecco:2,lopez-lopez:2016:gi-gecco,liou:2020:acm-taco,gu:2020:hipc,langdon:2015:gi-gecco,welton:2020:ics,sedgewick:1978:cacm,nguyen:2016:iukm,langdon:2014:gecco,langdon:2015:gecco,langdon:2017:gi-gecco,brownlee:2020:cec,bokhari:2018:mobiquitous,alcocer:2016:icpe,cody-kenny:2015:gi-gecco,li:2014:icse,katsaragakis:2020:iscas,devore-mcdonald:2020:oopsla,lopez-lopez:2018:gi-gecco,burles:2015:ssbse-1,yeboah-antwi:2017:gpem,fischer:2006:hpcmp-ugc,langdon:2015:tevc,doeraene:2016:oopsla,cody-kenny:2018:gi-icse,bartoli:2014:gecco,schulte:2014:asplos,honig:2014:trios,amidon:2015:hpec,an:2019:fse,bruce:2015:gecco,hung-cuong:2013:iccasa,khatchadourian:2019:icse,brownlee:2017:tetci,gutierrez:2014:icse,petke:2018:tse,stratis:2016:ase,sakamoto:2020:hpcasia,pereira:2016:greens-icse,haraldsson:2017:iscc,linderman:2010:cgo,brunie:2020:sc,petke:2014:eurogp}, \textbf{configuration~(62)}: \cite{luo:2009:emc,carvalho:2018:sac,hung:2009:icess,sandran:2011:socpros,desai:2009:iadcc,hernandez:2008:sp,jia:2013:qsic:1,hoste:2010:cgo,czako:2021:esa,dave:2009:lcpc,werner:2020:mlcad,varrette:2019:ppam,mesmay:2009:icml,desani:2016:pdp,ashouri:2016:taco,williams-king:2019:feast-ccs,chebolu:2016:icacci,ibrahim:2009:newcas,triantafyllis:2003:cgo,lima:2013:patmos,koo:2019:eurosec,bolat:2009:ipccc,wu:2015:gecco,bhattacharya:2012:sigmetrics,rasch:2021:taco,chen:2021:icse,cherian:2020:corr,ilham:2011:iceei,garciarena:2016:gi-gecco,dasilva:2021:cc,pan:2006:cgo,branco:2015:isci,grunwald:1993:pldi,sachan:2021:jsa,katsaragakis:2020:iscas,colucci:2020:corr,blot:2016:lion,azeemi:2006:iit,wang:2010:cgo,haneda:2006:ipdps,canales:2021:icpe-wip,fischer:2006:hpcmp-ugc,cooper:1999:lctes,akiba:2019:kdd,hutter:2009:jair,pinto:2022:tse,sandran:2012:cec,honig:2014:trios,abdusalam:2014:igcc,cavazos:2007:cgo,che:2005:hpcasia,brownlee:2017:tetci,hutter:2011:lion,qian:2020:ccs,jiang:2022:tr,canino:2018:fse,rahman:2012:cf,chan:1994:can,daud:2009:ijict,moyasebi:2020:gi-gecco,mcphee:2017:gecco-c_1,kulkarni:2006:tecs} \\

    \bottomrule
    \label{table:relevant_papers}
  \end{tabular}
\end{table*}

%% file: split_survey.tex
\section{Empirical Work on Non-Functional Properties of Software}
\label{section:survey}

\autoref{section:method} described the literature review process that resulted in a corpus of 386 papers related to improvement of non-functional improvement of software.
In this section, we examine these 386 paper in more detail, focusing on extracting information according to the following four criteria:
\begin{description}
\item[Research landscape.] For each paper we note the publication or release date, and the name and the type of venue in which it appeared.
\item[Fitness functions.] We note the non-functional properties targeted, and in the cases where multiple properties are reported, whether they have been actually used to produce improved software variants or simply measured at the end of the experiments. We also note how each of the non-functional properties was measured.
\item[Search Approaches.] We note both the type of approach used to generate software variants and the type of modifications applied to the original software.
\item[Benchmark.] We note the number and names of the software used in the empirical evaluation. We note if they were selected from an existing benchmark, as well as their size, programming language they were written in, their origin, and the platform on which they were run.
\end{description}

Unless explicitly stated otherwise (i.e., \autoref{figure:repo_by_fitness} and \autoref{figure:repo_by_fitness2}), all analyses in this section use the full corpus of 386 papers.

\subsection{Research Landscape}

\begin{figure}
  \centering
  \begin{minipage}[t]{0.48\textwidth}
    \input{figure/survey_years.tex}
  \end{minipage}\hfill
  \begin{minipage}[t]{0.48\textwidth}
    \centering
    \input{figure/survey_years_and_repo.tex}
  \end{minipage}
\end{figure}

\autoref{figure:survey_years} shows the publication year distribution of all 386 relevant papers;
almost all papers appeared after 1995, with a clear upward trend.
While it can be an artefact of the relevance sort of the online repositories, there is no doubt that work on software's non-functional properties is increasingly widespread.
Finally, most related work is published in conferences, although we note a fair number of workshop papers and a very high number of journal articles in 2021.

\autoref{figure:survey_years_and_repo} details the origin (i.e., preliminary search or digital library), by publication year, of all 386 corpus papers.
With the exception of two papers published in 2019, all relevant work obtained through Google Scholar appeared before 2011.
Conversely, ArXiV only yielded papers from 2019 onwards and ACM Digital Library from 2011 onwards.
However, both IEEE and Scopus yielded papers for every year since 1998.

\autoref{figure:venues} presents the venues in which the papers appear most frequently in (i.e., at least five papers published in a given venue).
With the notable exception of GECCO, an evolutionary computation conference with the highest figure of 20 related papers, venues are thematically split between applied computing (with CGO, ICS, PACT, IPDPS, and OOPSLA), for a total of 45 papers, and general software engineering (with ICSE, IEEE TSE, ASE, and ESEC/FSE), for a total of 28 papers.

\subsection{Fitness functions}

\begin{figure}
  \centering
  \begin{minipage}{0.48\textwidth}
    \input{figure/venues.tex}
  \end{minipage}\hfill
  \begin{minipage}{0.48\textwidth}
    \input{figure/survey_years_and_fitness.tex}
  \end{minipage}
\end{figure}

\autoref{figure:survey_years_and_fitness} shows the distribution of all 386 corpus papers over the years, according to the types of non-functional software property they target.
In order to increase readability, properties have been grouped thematically: e.g., ``Time'' includes mostly execution time (61\% of all papers), but also test or compilation time ($\approx$1\%), ``Memory'' encompasses both optimisation of the total amount of resource used, but also its actual usage during execution (e.g., minimising cache misses), and ``Size'' groups binary size on disk as well as source code size (e.g., in terms of lines of code).

By far the most frequent non-functional property targeted for improvement is time efficiency.
Time is optimised in 241 of the 386 corpus papers, followed by energy usage (71 papers), software size (69 papers) and memory usage (35 papers).
54 papers also targeted other types of properties, including mostly the quality of the output, the software's attack surface, or the overall writing quality of source code.
The remaining 54 papers targeted other properties, such as copyright or security issues, code complexity, or source code obfuscation.

\medskip
\centerline{
  \fbox{
    \parbox{0.97\linewidth}{
      \textbf{Answer to RQ1 (a):}
      Empirical work on optimisation of non-functional properties of software most frequently focuses on improvement of execution time.
      We found 241 out of 386 papers that describe such work.
      Code size follows with 69 papers (18 more as a secondary objective), while energy usage is targeted in 71 papers (and only 3 more as a secondary objective).
      Memory usage is considered in 35 papers, and in further 16 as a secondary objective.
      Work on improvement of other non-functional properties of software is more rare, with the next most frequently targeted property, `output quality', being improved in 25 papers, while `attack surface' being in 11.
    }
  }
}
\medskip

\begin{figure}
  \centering
  \begin{minipage}[t]{0.48\textwidth}
    \input{figure/survey_years_and_moo.tex}
  \end{minipage}\hfill
  \begin{minipage}[t]{0.48\textwidth}
    \centering
    \input{figure/fitness_by_fitness.tex}
  \end{minipage}
\end{figure}

Furthermore, as shown in \autoref{figure:survey_year_and_moo}, the authors of 132 papers considered more than one non-functional property.
We can distinguish 21 papers that proposed work that actively optimised multiple non-functional properties of software simultaneously, 57 papers that had a secondary focus on at least another property, and finally 54 other papers where authors simply reported on more than one property.
\autoref{figure:fitness_by_fitness} shows the pairs of fitness types considered in work simultaneously optimising at least two non-functional properties during search.
As expected, execution time is also the most popular fitness function considered.
We also note a few papers where both software's energy consumption and output solution quality was targeted.
Overall, whilst the number of such ``multi-objective papers'' is undeniably growing, work toward proper multi-objective improvement of non-functional properties is still quite rare.

\medskip
\centerline{
  \fbox{
    \parbox{0.97\linewidth}{
      \textbf{Answer to RQ1 (b):}
      There is little work on multi-objective improvement of non-functional properties of software (only 21 relevant papers found).
      In addition to being the most frequent objective (found 16 times), in most cases execution time is one of the targeted properties, often used as a trade-off to energy usage or solution quality (six times each).
    }}}
\medskip

\begin{figure}
  \begin{minipage}[t]{0.48\textwidth}
    \input{figure/repo_by_fitness.tex}
  \end{minipage}\hfill
  \begin{minipage}[t]{0.48\textwidth}
    \input{figure/repo_by_fitness2.tex}
  \end{minipage}
\end{figure}

\autoref{figure:repo_by_fitness} shows the correlation between the non-functional property keyword categories used in the systematic repository search and the actual primary fitness types targeted in the relevant papers.
Surprisingly, only ``energy'' and ``other'' keywords ---the latter including bloat- and size-related terms--- are effective in yielding thematically relevant papers.
Furthermore, ``quality'' and ``memory'' keywords are far more effective in yielding papers optimising software speed than their expected fitness, even more than ``time'' keywords.

Conversely, \autoref{figure:repo_by_fitness2} shows the correlation between the repositories and the primary targeted fitness.
First, as already pointed out, search through ArXiV and Google Scholar really wasn't effective.
Then, whilst yielding many more relevant results, both the ACM and the IEEE digital libraries show a bias towards work targeting running time improvements, the most frequent non-functional property overall.
Finally, surprisingly, Scopus does not seem to exhibit that strong a bias, also yielding high numbers of papers targeting size and energy concerns.

\subsection{Search Approaches}

Five main types of search approaches are distinguished.
(1) \emph{static} approaches in which decisions about software modifications are taken without recomputing fitness values; typically a single software variant is generated and compared to the original software.
(2) \emph{sampling} approaches, in which a given number of variants are generated and evaluated, the best variant being only determined at the end of the procedure.
We include in that category both random and systematic sampling, as well as exhaustive enumeration.
(3) \emph{exploratory} (non-evolutionary) approaches, in which multiple software variants are iteratively generated and evaluated in order to produce a final software variant.
Exploratory approaches differ from sampling approaches in that they are trajectory-based, intermediary variants being used to guide the search.
This category includes, for example, local search algorithms and greedy approaches, specifically excluding evolutionary approaches.
(4) \emph{evolutionary} approaches, that expand on exploratory approaches by using biology-inspired procedures such as, for example, genetic algorithms or genetic programming.
Finally, (5) \emph{manual} approaches in which the software is modified by hand.

\begin{figure}
  \begin{minipage}{0.48\textwidth}
    \input{figure/survey_years_and_search.tex}
  \end{minipage}\hfill
  \begin{minipage}{0.48\textwidth}
    \input{figure/survey_years_and_approach.tex}
  \end{minipage}
\end{figure}

\autoref{figure:years_and_search} shows the distribution of types of search approaches by year of publication.
Static approaches are the most frequent (204 papers), and constitute with manual (46 papers) and sampling (28 papers) approaches almost all publications until around 2005.
Both exploratory (52 papers) and evolutionary (71 papers) approaches start to appear around 2000, being more prevalent after 2005.
Finally, most evolutionary approaches appear after 2014.

\medskip
\centerline{
  \fbox{
    \parbox{0.97\linewidth}{
      \textbf{Answer to RQ1 (c):}
      Most of the work on improvement of non-functional properties of software use static approaches (53\%).
      However, for the past ten years evolutionary (18\%) and exploratory (14\%) approaches have been increasingly popular.
    }
  }
}
\medskip

Similarly, we distinguished between four types of software modifications.
(1)~\emph{Loop transformations}, encompassing, for example, loop merging, splitting, unrolling, polyhedral transformations of nested loops, but also many other modifications specifically targeting loops in source code.
(2)~\emph{Semantic}-based modifications, with, for example, template-based code generation or refactorings, which are meant to guarantee semantics preservation.
(3)~Potentially \emph{destructive} modifications that do not guarantee semantics preservation; they can preserve semantics, but do not come with such guarantees, typically leaving change acceptance to a code review process.
(4)~\emph{Configuration}-based modifications, in which parameter values at known decisions points are changed.

\autoref{figure:years_and_approach} shows the distribution of types of software modifications by year of publication.
The vast majority of work applies semantic modifications to the software at hand (222 papers).
Loop transformations (96 papers) are especially prevalent in compiler work.
Configuration (64 papers) has been regularly tackled from around 2005, while destructive modifications (77 papers) have been very popular in the last ten years.
Unsurprisingly, the use of software destructive modifications can be linked to the increased use of evolutionary search in software engineering and the introduction of genetic improvement~\cite{petke:2018:tevc}.

\medskip
\centerline{
  \fbox{
    \parbox{0.97\linewidth}{
      \textbf{Answer to RQ1 (d):}
      Most of the work on non-functional properties of software use semantic modifications (58\%).
      However, similarly to RQ1 (c), destructive source code changes (20\%) have been increasingly popular in the past ten years.
      Other types of modifications are loop transformations (25\%) and configuration tuning (17\%), steadily appearing in the literature for the past twenty years.
    }
  }
}
\medskip

\subsection{Benchmarks}

Empirical studies on improvement of non-functional program properties are evaluated on particular sets of instances to measure performance improvement.
Although almost all contain the name of the software that was improved, many lack details regarding the dataset they were evaluated on.

In what follows, we denote by \emph{benchmark} any pair of both a given software code to execute and the necessary input data required to make execution reproducible, so that performance across different environments can be fairly and reliably compared.
A \emph{benchmark suite} is a collection of benchmarks; whilst theoretically designed to be executed as a whole, many studies only consider subsets of one or more benchmark suites to solely focus on specifically chosen benchmarks.

\begin{figure}
  \begin{minipage}{0.48\textwidth}
    \input{figure/benchmark_sets.tex}
  \end{minipage}\hfill
  \begin{minipage}{0.48\textwidth}
    \input{figure/survey_years_and_softset.tex}
  \end{minipage}
\end{figure}

In 156 of the 386 papers on empirical work on improvement of non-functional properties of software authors reused existing benchmark suites (40.4\%).
\autoref{figure:benchmark_sets} shows the most commonly used benchmark suites.
For SPEC, we distinguished Java benchmark suites (SPECjbb, SPECjvm) from the C/C\texttt{++}/Fortran benchmark suites (mostly SPEC CPU when specified).
One issue we encountered was that many times authors simply mention ``SPEC'' without specifying which version of that benchmark suite was used, which specific software was used, or even which programming language was targeted.
Similarly, we grouped together the original and revised versions of both MediaBench (MediaBench II) and SPLASH (SPLASH-2).

SPEC is the most frequently used benchmark suite in work on non-functional property improvement of software.
This is partly due to the longevity of its various benchmark suites that are regularly updated (e.g., SPEC CPU has been revised five times since its inception).
\autoref{figure:year_and_benchmark} shows that despite a great number of benchmark suites being reused, and an increasing number of reuse taking place over the years, no particular suite appears to be prevalent.
One reason might be the increasing difficulty of compiling older software on newer systems, making benchmark suites quickly outdated, issue that SPEC may avoid by updating their benchmarks suites more frequently.

\medskip
\centerline{
  \fbox{
    \parbox{0.97\linewidth}{
      \textbf{Answer to RQ2 (a):}
      Whilst many software benchmark suites have been proposed, they are used in less than half (40.4\%) of the papers surveyed.
      SPEC, with its multiple types of benchmarks, is by far the most commonly reused benchmark suite.
      However, it was originally proposed for raw performance evaluation of hardware systems and may not be suitable for all non-functional property improvement purposes.
    }
  }
}
\medskip

\input{figure/benchmark_software.tex}

\autoref{figure:software_names} presents software most frequently targeted in the 386 papers, as well as when relevant, the benchmark suite it was explicitly taken from (we use ``--'' to indicate that no benchmark suite was specified in the paper).
Unsurprisingly, the most frequently targeted software come from the SPEC benchmark suite (23 of the 31 software targeted ten times or more).
Software not being part of benchmark suites are not frequently reused, with, for example, MiniSAT (found seven times) or LULESH (six times).

\medskip
\centerline{
  \fbox{
    \parbox{0.97\linewidth}{
      \textbf{Answer to RQ2 (b):}
      Unsurprisingly the most often targeted software are those originating from the SPEC benchmarks, including, for example, \texttt{bzip2}, \texttt{mcf}, and \texttt{gcc}.
    }
  }
}
\medskip

\begin{figure}
  \begin{minipage}[t]{0.48\textwidth}
    \input{figure/survey_years_and_language.tex}
  \end{minipage}\hfill
  \begin{minipage}[t]{0.48\textwidth}
    \input{figure/survey_years_and_soft_size.tex}
  \end{minipage}
\end{figure}

\autoref{figure:year_and_language} shows the distribution of the programming language of the targeted software relative to the publication year.
Targeted software are most frequently written in C or C\texttt{++}, with 54\% of all papers.
An additional 8\% of papers target GPU software (e.g., CUDA) essentially also written in C/C\texttt{++}.
Java follows with 16\% of papers, then Fortran (5\%, appearing in a single paper only since 2010) and Javascript (2\%).
Other languages (including Scala, Python, Erlang, Haskell...) only appear in 1\% or fewer papers.
Surprisingly, in 12\% of papers the programming language of the targeted software is not explicitly stated.

As for the type of the targeted software, 57\% of the papers include real-world software, usually freely available online, while 52\% considered toy examples specifically written for research purposes; in 10\% of the papers the origin of software was unclear.
Regarding papers with real-world software, which seem to be increasingly favoured, we find that 59\% is academic in origin, 47\% come from the open-source community, 6\% come from industry; 16\% is unknown.

For each paper we noted the approximate size of the software targeted within, with categories such as ``block'' (a few lines of code; 11\% of papers), ``function'' (a small number of functions; 16\% of papers), ``file'' (typically one or two files; 18\% of papers), and ``project'' (28\%) and finally ``application'' (41\%) for larger software with more files and a much larger code base.
In 36\% of papers the size of the targeted software was neither specified nor obvious.
\autoref{figure:year_and_size} details proportions by publication year.
Whilst results appear relatively constant, it can still be noted that the proportion of both small software (i.e., ``block'', ``function'', and ``file'') and software of unknown size seem to be slowly decreasing, in favour of large software.

Finally, we looked at the environment in which software was executed.
Unfortunately, in 50\% of the papers this was not specified and could not be inferred.
In 40\% of papers software was run on a Unix or Linux machine, 3\% on Windows, and 1\% on Mac specifically.
Additionally, 6\% of the papers considered software running on a mobile device (Android: 5\%, iPhone: 1\%).

\medskip
\centerline{
  \fbox{
    \parbox{0.97\linewidth}{
      \textbf{Answer to RQ3 (a):}
      The most frequently considered software for improvement of its non-functional behaviour is, statistically, a large real-world Linux application written in C or C\texttt{++}; this might be a direct consequence of the prevalence of work based on SPEC.
    }
  }
}

\medskip
\centerline{
  \fbox{
    \parbox{0.97\linewidth}{
      \textbf{Answer to RQ3 (b):}
      Benchmarks revealed in our survey are not representative of real-world software as a whole, with, for example, Java (60 papers, 15.5\%), Javascript (nine papers, 2.3\%), or Python (four papers, 1\%) software being extremely underrepresented in view of their actual popularity.
    }
  }
}

%% file: figure/survey_years.tex
  \centering
  \begin{tikzpicture}
    \begin{axis}[
        width=20em,
        height=16em,
        ybar stacked,
        ymin=0,
        xlabel={Publication year},
        ylabel={Unique corpus paper},
        grid,
        bar width=3pt,
        thick,
        cycle list/Set1,
        every axis plot/.append style={draw,fill,fill opacity=0.5},
        x tick label style={/pgf/number format/1000 sep=},
        try min ticks=5,
        legend pos=north west,
        reverse legend,
      ]
      \addplot coordinates {
        (1978,1)
        (1979,1)
        (1983,1)
        (1984,1)
        (1988,2)
        (1989,1)
        (1992,1)
        (1993,1)
        (1994,2)
        (1996,1)
        (1997,3)
        (1998,4)
        (1999,6)
        (2000,8)
        (2001,10)
        (2002,8)
        (2003,10)
        (2004,9)
        (2005,4)
        (2006,11)
        (2007,9)
        (2008,8)
        (2009,16)
        (2010,14)
        (2011,13)
        (2012,14)
        (2013,14)
        (2014,24)
        (2015,23)
        (2016,26)
        (2017,17)
        (2018,29)
        (2019,35)
        (2020,33)
        (2021,21)
        (2022,5)
      };
    \end{axis}
  \end{tikzpicture}
  \caption{Publication year distribution of all unique 386 corpus papers.}
  \label{figure:survey_years}

%% file: figure/survey_years_and_repo.tex
  \centering
  \begin{tikzpicture}
    \begin{axis}[
        width=20em,
        height=16em,
        ybar stacked,
        ymin=0,
        xlabel={Publication year},
        ylabel={Cumulative corpus paper},
        grid,
        bar width=3pt,
        thick,
        cycle list/Set1,
        every axis plot/.append style={draw,fill,fill opacity=0.5},
        x tick label style={/pgf/number format/1000 sep=},
        try min ticks=5,
        legend pos=north west,
        reverse legend,
      ]
      \addplot coordinates {
        (1977,0.000000)
        (1978,0.000000)
        (1979,0.000000)
        (1980,0.000000)
        (1981,0.000000)
        (1982,0.000000)
        (1983,0.000000)
        (1984,0.000000)
        (1985,0.000000)
        (1986,0.000000)
        (1987,0.000000)
        (1988,0.000000)
        (1989,0.000000)
        (1990,0.000000)
        (1991,0.000000)
        (1992,0.000000)
        (1993,0.000000)
        (1994,0.000000)
        (1995,0.000000)
        (1996,0.000000)
        (1997,2.000000)
        (1998,1.000000)
        (1999,2.000000)
        (2000,2.000000)
        (2001,7.000000)
        (2002,6.000000)
        (2003,5.000000)
        (2004,4.000000)
        (2005,2.000000)
        (2006,5.000000)
        (2007,3.000000)
        (2008,1.000000)
        (2009,7.000000)
        (2010,2.000000)
        (2011,1.000000)
        (2012,5.000000)
        (2013,3.000000)
        (2014,7.000000)
        (2015,7.000000)
        (2016,10.000000)
        (2017,4.000000)
        (2018,8.000000)
        (2019,14.000000)
        (2020,12.000000)
        (2021,5.000000)
        (2022,1.000000)
      };
      \addplot coordinates {
        (1977,0.000000)
        (1978,0.000000)
        (1979,0.000000)
        (1980,0.000000)
        (1981,0.000000)
        (1982,0.000000)
        (1983,0.000000)
        (1984,0.000000)
        (1985,0.000000)
        (1986,0.000000)
        (1987,0.000000)
        (1988,0.000000)
        (1989,0.000000)
        (1990,0.000000)
        (1991,0.000000)
        (1992,0.000000)
        (1993,0.000000)
        (1994,0.000000)
        (1995,0.000000)
        (1996,0.000000)
        (1997,0.000000)
        (1998,2.000000)
        (1999,2.000000)
        (2000,5.000000)
        (2001,3.000000)
        (2002,4.000000)
        (2003,2.000000)
        (2004,6.000000)
        (2005,1.000000)
        (2006,2.000000)
        (2007,4.000000)
        (2008,2.000000)
        (2009,6.000000)
        (2010,3.000000)
        (2011,3.000000)
        (2012,3.000000)
        (2013,5.000000)
        (2014,5.000000)
        (2015,6.000000)
        (2016,4.000000)
        (2017,5.000000)
        (2018,7.000000)
        (2019,8.000000)
        (2020,6.000000)
        (2021,6.000000)
        (2022,5.000000)
      };
      \addplot coordinates {
        (1977,0.000000)
        (1978,1.000000)
        (1979,1.000000)
        (1980,0.000000)
        (1981,0.000000)
        (1982,0.000000)
        (1983,0.000000)
        (1984,1.000000)
        (1985,0.000000)
        (1986,0.000000)
        (1987,0.000000)
        (1988,2.000000)
        (1989,0.000000)
        (1990,0.000000)
        (1991,0.000000)
        (1992,0.000000)
        (1993,0.000000)
        (1994,1.000000)
        (1995,0.000000)
        (1996,0.000000)
        (1997,0.000000)
        (1998,0.000000)
        (1999,0.000000)
        (2000,1.000000)
        (2001,1.000000)
        (2002,1.000000)
        (2003,0.000000)
        (2004,0.000000)
        (2005,1.000000)
        (2006,2.000000)
        (2007,0.000000)
        (2008,0.000000)
        (2009,1.000000)
        (2010,2.000000)
        (2011,5.000000)
        (2012,1.000000)
        (2013,2.000000)
        (2014,8.000000)
        (2015,10.000000)
        (2016,7.000000)
        (2017,7.000000)
        (2018,9.000000)
        (2019,14.000000)
        (2020,14.000000)
        (2021,8.000000)
        (2022,0.000000)
      };
      \addplot coordinates {
        (1977,0.000000)
        (1978,0.000000)
        (1979,0.000000)
        (1980,0.000000)
        (1981,0.000000)
        (1982,0.000000)
        (1983,0.000000)
        (1984,0.000000)
        (1985,0.000000)
        (1986,0.000000)
        (1987,0.000000)
        (1988,0.000000)
        (1989,0.000000)
        (1990,0.000000)
        (1991,0.000000)
        (1992,0.000000)
        (1993,0.000000)
        (1994,0.000000)
        (1995,0.000000)
        (1996,0.000000)
        (1997,0.000000)
        (1998,0.000000)
        (1999,0.000000)
        (2000,0.000000)
        (2001,0.000000)
        (2002,0.000000)
        (2003,0.000000)
        (2004,0.000000)
        (2005,0.000000)
        (2006,0.000000)
        (2007,0.000000)
        (2008,4.000000)
        (2009,4.000000)
        (2010,5.000000)
        (2011,4.000000)
        (2012,6.000000)
        (2013,5.000000)
        (2014,7.000000)
        (2015,9.000000)
        (2016,8.000000)
        (2017,5.000000)
        (2018,12.000000)
        (2019,6.000000)
        (2020,9.000000)
        (2021,3.000000)
        (2022,0.000000)
      };
      \addplot coordinates {
        (1977,0.000000)
        (1978,0.000000)
        (1979,0.000000)
        (1980,0.000000)
        (1981,0.000000)
        (1982,0.000000)
        (1983,1.000000)
        (1984,0.000000)
        (1985,0.000000)
        (1986,0.000000)
        (1987,0.000000)
        (1988,0.000000)
        (1989,1.000000)
        (1990,0.000000)
        (1991,0.000000)
        (1992,1.000000)
        (1993,1.000000)
        (1994,1.000000)
        (1995,0.000000)
        (1996,1.000000)
        (1997,1.000000)
        (1998,1.000000)
        (1999,3.000000)
        (2000,0.000000)
        (2001,1.000000)
        (2002,0.000000)
        (2003,3.000000)
        (2004,1.000000)
        (2005,0.000000)
        (2006,2.000000)
        (2007,2.000000)
        (2008,2.000000)
        (2009,0.000000)
        (2010,2.000000)
        (2011,1.000000)
        (2012,0.000000)
        (2013,0.000000)
        (2014,0.000000)
        (2015,0.000000)
        (2016,0.000000)
        (2017,0.000000)
        (2018,0.000000)
        (2019,1.000000)
        (2020,0.000000)
        (2021,0.000000)
        (2022,0.000000)
      };
      \addplot coordinates {
        (1977,0.000000)
        (1978,0.000000)
        (1979,0.000000)
        (1980,0.000000)
        (1981,0.000000)
        (1982,0.000000)
        (1983,0.000000)
        (1984,0.000000)
        (1985,0.000000)
        (1986,0.000000)
        (1987,0.000000)
        (1988,0.000000)
        (1989,0.000000)
        (1990,0.000000)
        (1991,0.000000)
        (1992,0.000000)
        (1993,0.000000)
        (1994,0.000000)
        (1995,0.000000)
        (1996,0.000000)
        (1997,0.000000)
        (1998,0.000000)
        (1999,0.000000)
        (2000,0.000000)
        (2001,0.000000)
        (2002,0.000000)
        (2003,0.000000)
        (2004,0.000000)
        (2005,0.000000)
        (2006,0.000000)
        (2007,0.000000)
        (2008,0.000000)
        (2009,0.000000)
        (2010,0.000000)
        (2011,0.000000)
        (2012,0.000000)
        (2013,0.000000)
        (2014,0.000000)
        (2015,0.000000)
        (2016,0.000000)
        (2017,0.000000)
        (2018,0.000000)
        (2019,3.000000)
        (2020,4.000000)
        (2021,3.000000)
        (2022,0.000000)
      };

      \legend{Scopus (126),IEEE (105),Manual search (100),ACM (87),Google Scholar (26),ArXiV (10)}
    \end{axis}
  \end{tikzpicture}
  \caption{Publication year distribution of all 386 corpus papers, according to origin (preliminary manual search or digital library). Papers found multiple times are counted multiple times.}
  \label{figure:survey_years_and_repo}

%% file: figure/venues.tex
  \centering
  \begin{tikzpicture}
    \begin{axis}[
        width=17em,
        xbar,
        xmin=0,
        xlabel={Number of papers by venue},
        ylabel={Publication venue},
        ytick={0,1,2,3,4,5,6,7,8,9,10,11,12,13,14,15,16},
        yticklabels={SC,ASPLOS,ASE,IPDPS,IEEE TSE,CCS,PACT,OOPSLA,ESEC/FSE,PLDI,ICS,IEEE TPDS,ArXiV,ICSE,CGO,GECCO},
        yticklabel style={font=\small},
        nodes near coords={\pgfkeys{/pgf/fpu}\pgfmathparse{\pgfplotspointmeta}\pgfmathprintnumber[fixed,precision=0]{\pgfmathresult}},
        nodes near coords align={horizontal},
        bar shift=0pt,
        enlarge x limits={upper, value=0.15},
        enlarge y limits={true, abs value=0.8},
        bar width=0.6em,
        y=0.8em,
        legend pos={south east},
        thick,
      ]
      \addplot coordinates {
        (5,0) 
        (5,1) 
        (5,2) 
        (6,3) 
        (6,5) 
        (7,6) 
        (7,7) 
        (7,8) 
        (9,9) 
        (9,10) 
        (13,13) 
        (18,14) 
        (20,15) 
      };
      \addplot coordinates {
        (6,4) 
        (10,11) 
      };
      \addplot coordinates {
        (10,12) 
      };
      \legend{Conference,Journal,ArXiV};
    \end{axis}
  \end{tikzpicture}
  \caption{Most frequent publication venues ($\ge$5) across all 386 corpus papers.}
  \label{figure:venues}

%% file: figure/survey_years_and_fitness.tex
  \centering
  \begin{tikzpicture}
    \begin{axis}[
        width=20em,
        height=16em,
        ybar stacked,
        ymin=0,
        xlabel={Publication year},
        ylabel={Cumulative corpus paper},
        grid,
        bar width=3pt,
        thick,
        cycle list/Set1,
        every axis plot/.append style={fill,fill opacity=0.5},
        x tick label style={/pgf/number format/1000 sep=},
        try min ticks=5,
        legend pos=north west,
        reverse legend,
      ]
      \addplot coordinates {
        (1977,0.000000)
        (1978,1.000000)
        (1979,1.000000)
        (1980,0.000000)
        (1981,0.000000)
        (1982,0.000000)
        (1983,1.000000)
        (1984,1.000000)
        (1985,0.000000)
        (1986,0.000000)
        (1987,0.000000)
        (1988,1.000000)
        (1989,0.000000)
        (1990,0.000000)
        (1991,0.000000)
        (1992,1.000000)
        (1993,0.000000)
        (1994,2.000000)
        (1995,0.000000)
        (1996,1.000000)
        (1997,1.000000)
        (1998,3.000000)
        (1999,4.000000)
        (2000,8.000000)
        (2001,8.000000)
        (2002,4.000000)
        (2003,5.000000)
        (2004,4.000000)
        (2005,3.000000)
        (2006,7.000000)
        (2007,6.000000)
        (2008,6.000000)
        (2009,10.000000)
        (2010,13.000000)
        (2011,10.000000)
        (2012,10.000000)
        (2013,9.000000)
        (2014,16.000000)
        (2015,16.000000)
        (2016,15.000000)
        (2017,8.000000)
        (2018,20.000000)
        (2019,15.000000)
        (2020,14.000000)
        (2021,15.000000)
        (2022,2.000000)
      };
      \addplot coordinates {
        (1977,0.000000)
        (1978,0.000000)
        (1979,0.000000)
        (1980,0.000000)
        (1981,0.000000)
        (1982,0.000000)
        (1983,0.000000)
        (1984,0.000000)
        (1985,0.000000)
        (1986,0.000000)
        (1987,0.000000)
        (1988,0.000000)
        (1989,0.000000)
        (1990,0.000000)
        (1991,0.000000)
        (1992,0.000000)
        (1993,0.000000)
        (1994,0.000000)
        (1995,0.000000)
        (1996,0.000000)
        (1997,0.000000)
        (1998,1.000000)
        (1999,0.000000)
        (2000,1.000000)
        (2001,2.000000)
        (2002,1.000000)
        (2003,1.000000)
        (2004,2.000000)
        (2005,0.000000)
        (2006,3.000000)
        (2007,3.000000)
        (2008,0.000000)
        (2009,5.000000)
        (2010,0.000000)
        (2011,2.000000)
        (2012,2.000000)
        (2013,2.000000)
        (2014,7.000000)
        (2015,6.000000)
        (2016,7.000000)
        (2017,5.000000)
        (2018,7.000000)
        (2019,7.000000)
        (2020,4.000000)
        (2021,2.000000)
        (2022,1.000000)
      };
      \addplot coordinates {
        (1977,0.000000)
        (1978,0.000000)
        (1979,0.000000)
        (1980,0.000000)
        (1981,0.000000)
        (1982,0.000000)
        (1983,1.000000)
        (1984,0.000000)
        (1985,0.000000)
        (1986,0.000000)
        (1987,0.000000)
        (1988,0.000000)
        (1989,0.000000)
        (1990,0.000000)
        (1991,0.000000)
        (1992,0.000000)
        (1993,0.000000)
        (1994,0.000000)
        (1995,0.000000)
        (1996,0.000000)
        (1997,2.000000)
        (1998,0.000000)
        (1999,2.000000)
        (2000,0.000000)
        (2001,2.000000)
        (2002,5.000000)
        (2003,3.000000)
        (2004,2.000000)
        (2005,1.000000)
        (2006,4.000000)
        (2007,1.000000)
        (2008,2.000000)
        (2009,2.000000)
        (2010,2.000000)
        (2011,0.000000)
        (2012,3.000000)
        (2013,1.000000)
        (2014,3.000000)
        (2015,3.000000)
        (2016,3.000000)
        (2017,3.000000)
        (2018,3.000000)
        (2019,7.000000)
        (2020,7.000000)
        (2021,4.000000)
        (2022,3.000000)
      };
      \addplot coordinates {
        (1977,0.000000)
        (1978,0.000000)
        (1979,0.000000)
        (1980,0.000000)
        (1981,0.000000)
        (1982,0.000000)
        (1983,0.000000)
        (1984,0.000000)
        (1985,0.000000)
        (1986,0.000000)
        (1987,0.000000)
        (1988,1.000000)
        (1989,0.000000)
        (1990,0.000000)
        (1991,0.000000)
        (1992,0.000000)
        (1993,0.000000)
        (1994,0.000000)
        (1995,0.000000)
        (1996,0.000000)
        (1997,0.000000)
        (1998,0.000000)
        (1999,0.000000)
        (2000,0.000000)
        (2001,0.000000)
        (2002,1.000000)
        (2003,1.000000)
        (2004,0.000000)
        (2005,1.000000)
        (2006,0.000000)
        (2007,0.000000)
        (2008,0.000000)
        (2009,0.000000)
        (2010,0.000000)
        (2011,1.000000)
        (2012,1.000000)
        (2013,3.000000)
        (2014,3.000000)
        (2015,1.000000)
        (2016,2.000000)
        (2017,5.000000)
        (2018,5.000000)
        (2019,14.000000)
        (2020,13.000000)
        (2021,2.000000)
        (2022,0.000000)
      };
      \addplot coordinates {
        (1977,0.000000)
        (1978,0.000000)
        (1979,0.000000)
        (1980,0.000000)
        (1981,0.000000)
        (1982,0.000000)
        (1983,0.000000)
        (1984,0.000000)
        (1985,0.000000)
        (1986,0.000000)
        (1987,0.000000)
        (1988,0.000000)
        (1989,1.000000)
        (1990,0.000000)
        (1991,0.000000)
        (1992,1.000000)
        (1993,1.000000)
        (1994,0.000000)
        (1995,0.000000)
        (1996,0.000000)
        (1997,0.000000)
        (1998,1.000000)
        (1999,1.000000)
        (2000,0.000000)
        (2001,2.000000)
        (2002,1.000000)
        (2003,1.000000)
        (2004,2.000000)
        (2005,0.000000)
        (2006,0.000000)
        (2007,1.000000)
        (2008,0.000000)
        (2009,0.000000)
        (2010,0.000000)
        (2011,2.000000)
        (2012,3.000000)
        (2013,2.000000)
        (2014,2.000000)
        (2015,3.000000)
        (2016,3.000000)
        (2017,0.000000)
        (2018,3.000000)
        (2019,2.000000)
        (2020,2.000000)
        (2021,1.000000)
        (2022,0.000000)
      };

      \legend{Time (241),Energy (71),Size (69),Other (54),Memory (35)}
    \end{axis}
  \end{tikzpicture}
  \caption{Publication year distribution of all 386 corpus papers, according to the types of non-functional software property targeted. Papers targeting multiple property types are counted multiple times.}
  \label{figure:survey_years_and_fitness}


%% file: figure/survey_years_and_moo.tex
  \centering
  \begin{tikzpicture}
    \begin{axis}[
        width=20em,
        height=16em,
        ybar stacked,
        ymin=0,
        xlabel={Publication year},
        ylabel={Unique corpus paper},
        grid,
        bar width=3pt,
        thick,
        cycle list/Set1,
        every axis plot/.append style={fill,fill opacity=0.5},
        x tick label style={/pgf/number format/1000 sep=},
        legend pos=north west,
        reverse legend,
      ]
      \addplot coordinates {
        (1978,0.000000)
        (1979,0.000000)
        (1980,0.000000)
        (1981,0.000000)
        (1982,0.000000)
        (1983,1.000000)
        (1984,0.000000)
        (1985,0.000000)
        (1986,0.000000)
        (1987,0.000000)
        (1988,0.000000)
        (1989,0.000000)
        (1990,0.000000)
        (1991,0.000000)
        (1992,1.000000)
        (1993,0.000000)
        (1994,0.000000)
        (1995,0.000000)
        (1996,0.000000)
        (1997,0.000000)
        (1998,1.000000)
        (1999,1.000000)
        (2000,1.000000)
        (2001,4.000000)
        (2002,3.000000)
        (2003,1.000000)
        (2004,1.000000)
        (2005,0.000000)
        (2006,1.000000)
        (2007,2.000000)
        (2008,0.000000)
        (2009,1.000000)
        (2010,1.000000)
        (2011,0.000000)
        (2012,5.000000)
        (2013,1.000000)
        (2014,5.000000)
        (2015,4.000000)
        (2016,2.000000)
        (2017,3.000000)
        (2018,5.000000)
        (2019,6.000000)
        (2020,5.000000)
        (2021,2.000000)
        (2022,0.000000)
      };
      \addplot coordinates {
        (1978,0.000000)
        (1979,1.000000)
        (1980,0.000000)
        (1981,0.000000)
        (1982,0.000000)
        (1983,0.000000)
        (1984,0.000000)
        (1985,0.000000)
        (1986,0.000000)
        (1987,0.000000)
        (1988,0.000000)
        (1989,0.000000)
        (1990,0.000000)
        (1991,0.000000)
        (1992,0.000000)
        (1993,1.000000)
        (1994,0.000000)
        (1995,0.000000)
        (1996,1.000000)
        (1997,0.000000)
        (1998,1.000000)
        (1999,0.000000)
        (2000,2.000000)
        (2001,1.000000)
        (2002,0.000000)
        (2003,2.000000)
        (2004,3.000000)
        (2005,0.000000)
        (2006,1.000000)
        (2007,0.000000)
        (2008,0.000000)
        (2009,3.000000)
        (2010,0.000000)
        (2011,4.000000)
        (2012,1.000000)
        (2013,5.000000)
        (2014,2.000000)
        (2015,5.000000)
        (2016,2.000000)
        (2017,1.000000)
        (2018,4.000000)
        (2019,7.000000)
        (2020,3.000000)
        (2021,2.000000)
        (2022,2.000000)
      };
      \addplot coordinates {
        (1978,0.000000)
        (1979,0.000000)
        (1980,0.000000)
        (1981,0.000000)
        (1982,0.000000)
        (1983,0.000000)
        (1984,0.000000)
        (1985,0.000000)
        (1986,0.000000)
        (1987,0.000000)
        (1988,0.000000)
        (1989,0.000000)
        (1990,0.000000)
        (1991,0.000000)
        (1992,0.000000)
        (1993,0.000000)
        (1994,0.000000)
        (1995,0.000000)
        (1996,0.000000)
        (1997,0.000000)
        (1998,0.000000)
        (1999,0.000000)
        (2000,0.000000)
        (2001,0.000000)
        (2002,0.000000)
        (2003,0.000000)
        (2004,0.000000)
        (2005,1.000000)
        (2006,2.000000)
        (2007,0.000000)
        (2008,0.000000)
        (2009,0.000000)
        (2010,1.000000)
        (2011,2.000000)
        (2012,0.000000)
        (2013,1.000000)
        (2014,1.000000)
        (2015,1.000000)
        (2016,1.000000)
        (2017,1.000000)
        (2018,3.000000)
        (2019,3.000000)
        (2020,2.000000)
        (2021,1.000000)
        (2022,1.000000)
      };

      \legend{Secondary focus (57),Reports multiple (54),Joint optimisation (21)}
    \end{axis}
  \end{tikzpicture}
  \caption{Publication year of all 132 corpus papers that consider more than one non-functional property, according to their multi-objective philosophy.}
  \label{figure:survey_year_and_moo}


%% file: figure/fitness_by_fitness.tex
  \centering
  \begin{tikzpicture}
    \begin{axis}[
        width=18em,
        xlabel={Property optimised},
        ylabel={Property optimised},
        xtick=data,
        ytick=data,
        symbolic x coords={Time,Other,Energy,Memory,Size},
        symbolic y coords={Time,Other,Energy,Memory,Size},
        grid,
        thick,
        cycle list/Set1,
        every tick label/.append style={font=\small},
      ]
      \addplot[%
          scatter=true,
          only marks,
          mark=*,
          point meta=explicit,
          visualization depends on = {\thisrow{v} \as \perpointmarksize},
          scatter/@pre marker code/.append style={/tikz/mark size=\perpointmarksize},
      ] table [x=x,y=y,meta index=2,col sep=semicolon,trim cells] {
x ; y ; v
Time ; Time ; 2.000000
Other ; Time ; 6.000000
Energy ; Time ; 6.000000
Memory ; Time ; 3.000000
Size ; Time ; 2.000000
Time ; Other ; 6.000000
Other ; Other ; 1.000000
Energy ; Other ; 4.000000
Memory ; Other ; 1.000000
Size ; Other ; 0.000000
Time ; Energy ; 6.000000
Other ; Energy ; 4.000000
Energy ; Energy ; 0.000000
Memory ; Energy ; 0.000000
Size ; Energy ; 0.000000
Time ; Memory ; 3.000000
Other ; Memory ; 1.000000
Energy ; Memory ; 0.000000
Memory ; Memory ; 0.000000
Size ; Memory ; 0.000000
Time ; Size ; 2.000000
Other ; Size ; 0.000000
Energy ; Size ; 0.000000
Memory ; Size ; 0.000000
Size ; Size ; 0.000000
};
    \end{axis}
  \end{tikzpicture}
  \caption{Correlation between non-functional properties considered in multi-objective work.}
  \label{figure:fitness_by_fitness}

%% file: figure/repo_by_fitness.tex
  \centering
  \begin{tikzpicture}
    \begin{axis}[
        width=18em,
        xlabel={Property optimised},
        ylabel={Property keyword category},
        xtick=data,
        ytick=data,
        symbolic x coords={Time,Size,Energy,Other,Memory},
        symbolic y coords={{Other},{Memory},{Quality},{Energy},{Time}},
        grid,
        thick,
        cycle list/Set1,
        every tick label/.append style={font=\small},
      ]
      \addplot[%
          scatter=true,
          only marks,
          mark=*,
          point meta=explicit,
          visualization depends on = {0.10*\thisrow{v} \as \perpointmarksize},
          scatter/@pre marker code/.append style={/tikz/mark size=\perpointmarksize},
      ] table [x=x,y=y,meta index=2,col sep=semicolon,trim cells] {
x ; y ; v
Time ; {Other} ; 48.000000
Size ; {Other} ; 80.000000
Energy ; {Other} ; 11.000000
Other ; {Other} ; 28.000000
Memory ; {Other} ; 5.000000
Time ; {Memory} ; 80.000000
Size ; {Memory} ; 17.000000
Energy ; {Memory} ; 3.000000
Other ; {Memory} ; 3.000000
Memory ; {Memory} ; 24.000000
Time ; {Quality} ; 79.000000
Size ; {Quality} ; 16.000000
Energy ; {Quality} ; 14.000000
Other ; {Quality} ; 8.000000
Memory ; {Quality} ; 7.000000
Time ; {Energy} ; 33.000000
Size ; {Energy} ; 6.000000
Energy ; {Energy} ; 72.000000
Other ; {Energy} ; 7.000000
Memory ; {Energy} ; 5.000000
Time ; {Time} ; 48.000000
Size ; {Time} ; 6.000000
Energy ; {Time} ; 3.000000
Other ; {Time} ; 6.000000
Memory ; {Time} ; 4.000000
};
    \end{axis}
  \end{tikzpicture}
  \caption{Correlation between non-functional property keyword category and actual targeted fitness (305 repository papers).}
  \label{figure:repo_by_fitness}

%% file: figure/repo_by_fitness2.tex
  \centering
  \begin{tikzpicture}
    \begin{axis}[
        width=18em,
        xlabel={Property optimised},
        ylabel={Digital library},
        xtick=data,
        ytick=data,
        symbolic x coords={Time,Size,Energy,Other,Memory},
        symbolic y coords={{Scopus},{IEEE},{ACM},{Scholar},{ArXiV}},
        grid,
        thick,
        cycle list/Set1,
        every tick label/.append style={font=\small},
      ]
      \addplot[%
          scatter=true,
          only marks,
          mark=*,
          point meta=explicit,
          visualization depends on = {0.10*\thisrow{v} \as \perpointmarksize},
          scatter/@pre marker code/.append style={/tikz/mark size=\perpointmarksize},
      ] table [x=x,y=y,meta index=2,col sep=semicolon,trim cells] {
x ; y ; v
Time ; {Scopus} ; 70.000000
Size ; {Scopus} ; 59.000000
Energy ; {Scopus} ; 46.000000
Other ; {Scopus} ; 21.000000
Memory ; {Scopus} ; 14.000000
Time ; {IEEE} ; 104.000000
Size ; {IEEE} ; 32.000000
Energy ; {IEEE} ; 32.000000
Other ; {IEEE} ; 12.000000
Memory ; {IEEE} ; 19.000000
Time ; {ACM} ; 80.000000
Size ; {ACM} ; 31.000000
Energy ; {ACM} ; 21.000000
Other ; {ACM} ; 12.000000
Memory ; {ACM} ; 5.000000
Time ; {Scholar} ; 26.000000
Size ; {Scholar} ; 2.000000
Energy ; {Scholar} ; 0.000000
Other ; {Scholar} ; 2.000000
Memory ; {Scholar} ; 5.000000
Time ; {ArXiV} ; 8.000000
Size ; {ArXiV} ; 1.000000
Energy ; {ArXiV} ; 4.000000
Other ; {ArXiV} ; 5.000000
Memory ; {ArXiV} ; 2.000000
};
    \end{axis}
  \end{tikzpicture}
  \caption{Correlation between digital library of origin and targeted fitness (305 repository papers).}
  \label{figure:repo_by_fitness2}

%% file: figure/survey_years_and_search.tex
  \centering
  \begin{tikzpicture}
    \begin{axis}[
        width=20em,
        height=16em,
        ybar stacked,
        ymin=0,
        xlabel={Publication year},
        ylabel={Cumulative corpus paper},
        grid,
        bar width=3pt,
        thick,
        cycle list/Set1,
        every axis plot/.append style={draw,fill,fill opacity=0.5},
        x tick label style={/pgf/number format/1000 sep=},
        legend pos=north west,
        reverse legend,
      ]
      \addplot coordinates {
        (1977,0.000000)
        (1978,0.000000)
        (1979,0.000000)
        (1980,0.000000)
        (1981,0.000000)
        (1982,0.000000)
        (1983,1.000000)
        (1984,0.000000)
        (1985,0.000000)
        (1986,0.000000)
        (1987,0.000000)
        (1988,2.000000)
        (1989,0.000000)
        (1990,0.000000)
        (1991,0.000000)
        (1992,0.000000)
        (1993,0.000000)
        (1994,0.000000)
        (1995,0.000000)
        (1996,1.000000)
        (1997,2.000000)
        (1998,4.000000)
        (1999,4.000000)
        (2000,6.000000)
        (2001,8.000000)
        (2002,6.000000)
        (2003,7.000000)
        (2004,6.000000)
        (2005,2.000000)
        (2006,5.000000)
        (2007,6.000000)
        (2008,6.000000)
        (2009,5.000000)
        (2010,10.000000)
        (2011,6.000000)
        (2012,7.000000)
        (2013,8.000000)
        (2014,12.000000)
        (2015,12.000000)
        (2016,10.000000)
        (2017,7.000000)
        (2018,16.000000)
        (2019,18.000000)
        (2020,15.000000)
        (2021,9.000000)
        (2022,3.000000)
      };
      \addplot coordinates {
        (1977,0.000000)
        (1978,0.000000)
        (1979,0.000000)
        (1980,0.000000)
        (1981,0.000000)
        (1982,0.000000)
        (1983,0.000000)
        (1984,0.000000)
        (1985,0.000000)
        (1986,0.000000)
        (1987,0.000000)
        (1988,0.000000)
        (1989,0.000000)
        (1990,0.000000)
        (1991,0.000000)
        (1992,0.000000)
        (1993,0.000000)
        (1994,0.000000)
        (1995,0.000000)
        (1996,0.000000)
        (1997,0.000000)
        (1998,0.000000)
        (1999,1.000000)
        (2000,0.000000)
        (2001,0.000000)
        (2002,1.000000)
        (2003,0.000000)
        (2004,0.000000)
        (2005,0.000000)
        (2006,3.000000)
        (2007,0.000000)
        (2008,0.000000)
        (2009,1.000000)
        (2010,2.000000)
        (2011,2.000000)
        (2012,1.000000)
        (2013,2.000000)
        (2014,4.000000)
        (2015,8.000000)
        (2016,6.000000)
        (2017,9.000000)
        (2018,7.000000)
        (2019,11.000000)
        (2020,3.000000)
        (2021,7.000000)
        (2022,1.000000)
      };
      \addplot coordinates {
        (1977,0.000000)
        (1978,0.000000)
        (1979,0.000000)
        (1980,0.000000)
        (1981,0.000000)
        (1982,0.000000)
        (1983,0.000000)
        (1984,0.000000)
        (1985,0.000000)
        (1986,0.000000)
        (1987,0.000000)
        (1988,0.000000)
        (1989,0.000000)
        (1990,0.000000)
        (1991,0.000000)
        (1992,0.000000)
        (1993,0.000000)
        (1994,0.000000)
        (1995,0.000000)
        (1996,0.000000)
        (1997,0.000000)
        (1998,0.000000)
        (1999,1.000000)
        (2000,0.000000)
        (2001,0.000000)
        (2002,1.000000)
        (2003,1.000000)
        (2004,1.000000)
        (2005,0.000000)
        (2006,1.000000)
        (2007,4.000000)
        (2008,1.000000)
        (2009,7.000000)
        (2010,1.000000)
        (2011,2.000000)
        (2012,1.000000)
        (2013,1.000000)
        (2014,3.000000)
        (2015,0.000000)
        (2016,3.000000)
        (2017,1.000000)
        (2018,2.000000)
        (2019,3.000000)
        (2020,10.000000)
        (2021,6.000000)
        (2022,2.000000)
      };
      \addplot coordinates {
        (1977,0.000000)
        (1978,1.000000)
        (1979,1.000000)
        (1980,0.000000)
        (1981,0.000000)
        (1982,0.000000)
        (1983,0.000000)
        (1984,1.000000)
        (1985,0.000000)
        (1986,0.000000)
        (1987,0.000000)
        (1988,0.000000)
        (1989,0.000000)
        (1990,0.000000)
        (1991,0.000000)
        (1992,1.000000)
        (1993,0.000000)
        (1994,1.000000)
        (1995,0.000000)
        (1996,0.000000)
        (1997,1.000000)
        (1998,0.000000)
        (1999,0.000000)
        (2000,1.000000)
        (2001,1.000000)
        (2002,0.000000)
        (2003,1.000000)
        (2004,0.000000)
        (2005,1.000000)
        (2006,1.000000)
        (2007,1.000000)
        (2008,0.000000)
        (2009,2.000000)
        (2010,1.000000)
        (2011,2.000000)
        (2012,4.000000)
        (2013,1.000000)
        (2014,2.000000)
        (2015,1.000000)
        (2016,5.000000)
        (2017,2.000000)
        (2018,5.000000)
        (2019,4.000000)
        (2020,5.000000)
        (2021,0.000000)
        (2022,0.000000)
      };
      \addplot coordinates {
        (1977,0.000000)
        (1978,0.000000)
        (1979,0.000000)
        (1980,0.000000)
        (1981,0.000000)
        (1982,0.000000)
        (1983,0.000000)
        (1984,0.000000)
        (1985,0.000000)
        (1986,0.000000)
        (1987,0.000000)
        (1988,0.000000)
        (1989,1.000000)
        (1990,0.000000)
        (1991,0.000000)
        (1992,0.000000)
        (1993,1.000000)
        (1994,1.000000)
        (1995,0.000000)
        (1996,0.000000)
        (1997,0.000000)
        (1998,0.000000)
        (1999,0.000000)
        (2000,1.000000)
        (2001,1.000000)
        (2002,0.000000)
        (2003,1.000000)
        (2004,2.000000)
        (2005,1.000000)
        (2006,1.000000)
        (2007,0.000000)
        (2008,1.000000)
        (2009,1.000000)
        (2010,0.000000)
        (2011,1.000000)
        (2012,2.000000)
        (2013,2.000000)
        (2014,3.000000)
        (2015,2.000000)
        (2016,2.000000)
        (2017,0.000000)
        (2018,1.000000)
        (2019,0.000000)
        (2020,1.000000)
        (2021,0.000000)
        (2022,1.000000)
      };

      \legend{Static (204),Evolutionary (69),Exploratory (52),Manual (46),Sampling (27)}
    \end{axis}
  \end{tikzpicture}
  \caption{Publication year distribution of all 386 corpus papers, according to the types of search approaches. Papers using multiple search approaches are counted multiple times.}
  \label{figure:years_and_search}

%% file: figure/survey_years_and_approach.tex
  \centering
  \begin{tikzpicture}
    \begin{axis}[
        width=20em,
        height=16em,
        ybar stacked,
        ymin=0,
        xlabel={Publication year},
        ylabel={Cumulative corpus paper},
        grid,
        bar width=3pt,
        thick,
        cycle list/Set1,
        every axis plot/.append style={draw,fill,fill opacity=0.5},
        x tick label style={/pgf/number format/1000 sep=},
        legend pos=north west,
        reverse legend,
      ]
      \addplot coordinates {
        (1977,0.000000)
        (1978,1.000000)
        (1979,1.000000)
        (1980,0.000000)
        (1981,0.000000)
        (1982,0.000000)
        (1983,1.000000)
        (1984,1.000000)
        (1985,0.000000)
        (1986,0.000000)
        (1987,0.000000)
        (1988,3.000000)
        (1989,2.000000)
        (1990,0.000000)
        (1991,0.000000)
        (1992,1.000000)
        (1993,0.000000)
        (1994,0.000000)
        (1995,0.000000)
        (1996,1.000000)
        (1997,3.000000)
        (1998,4.000000)
        (1999,4.000000)
        (2000,4.000000)
        (2001,7.000000)
        (2002,4.000000)
        (2003,8.000000)
        (2004,4.000000)
        (2005,4.000000)
        (2006,5.000000)
        (2007,6.000000)
        (2008,3.000000)
        (2009,9.000000)
        (2010,7.000000)
        (2011,7.000000)
        (2012,12.000000)
        (2013,7.000000)
        (2014,14.000000)
        (2015,10.000000)
        (2016,16.000000)
        (2017,10.000000)
        (2018,16.000000)
        (2019,18.000000)
        (2020,17.000000)
        (2021,8.000000)
        (2022,5.000000)
      };
      \addplot coordinates {
        (1977,0.000000)
        (1978,1.000000)
        (1979,0.000000)
        (1980,0.000000)
        (1981,0.000000)
        (1982,0.000000)
        (1983,0.000000)
        (1984,0.000000)
        (1985,0.000000)
        (1986,0.000000)
        (1987,0.000000)
        (1988,0.000000)
        (1989,0.000000)
        (1990,0.000000)
        (1991,0.000000)
        (1992,0.000000)
        (1993,0.000000)
        (1994,1.000000)
        (1995,0.000000)
        (1996,0.000000)
        (1997,0.000000)
        (1998,1.000000)
        (1999,3.000000)
        (2000,5.000000)
        (2001,4.000000)
        (2002,4.000000)
        (2003,3.000000)
        (2004,5.000000)
        (2005,0.000000)
        (2006,2.000000)
        (2007,2.000000)
        (2008,4.000000)
        (2009,3.000000)
        (2010,6.000000)
        (2011,3.000000)
        (2012,3.000000)
        (2013,4.000000)
        (2014,6.000000)
        (2015,6.000000)
        (2016,4.000000)
        (2017,0.000000)
        (2018,10.000000)
        (2019,5.000000)
        (2020,3.000000)
        (2021,5.000000)
        (2022,3.000000)
      };
      \addplot coordinates {
        (1977,0.000000)
        (1978,1.000000)
        (1979,0.000000)
        (1980,0.000000)
        (1981,0.000000)
        (1982,0.000000)
        (1983,0.000000)
        (1984,0.000000)
        (1985,0.000000)
        (1986,0.000000)
        (1987,0.000000)
        (1988,0.000000)
        (1989,0.000000)
        (1990,0.000000)
        (1991,0.000000)
        (1992,0.000000)
        (1993,0.000000)
        (1994,0.000000)
        (1995,0.000000)
        (1996,0.000000)
        (1997,0.000000)
        (1998,0.000000)
        (1999,0.000000)
        (2000,0.000000)
        (2001,0.000000)
        (2002,1.000000)
        (2003,0.000000)
        (2004,0.000000)
        (2005,0.000000)
        (2006,1.000000)
        (2007,0.000000)
        (2008,0.000000)
        (2009,0.000000)
        (2010,2.000000)
        (2011,1.000000)
        (2012,3.000000)
        (2013,5.000000)
        (2014,7.000000)
        (2015,10.000000)
        (2016,10.000000)
        (2017,7.000000)
        (2018,6.000000)
        (2019,11.000000)
        (2020,11.000000)
        (2021,2.000000)
        (2022,0.000000)
      };
      \addplot coordinates {
        (1977,0.000000)
        (1978,0.000000)
        (1979,0.000000)
        (1980,0.000000)
        (1981,0.000000)
        (1982,0.000000)
        (1983,0.000000)
        (1984,0.000000)
        (1985,0.000000)
        (1986,0.000000)
        (1987,0.000000)
        (1988,0.000000)
        (1989,0.000000)
        (1990,0.000000)
        (1991,0.000000)
        (1992,0.000000)
        (1993,1.000000)
        (1994,1.000000)
        (1995,0.000000)
        (1996,0.000000)
        (1997,0.000000)
        (1998,0.000000)
        (1999,1.000000)
        (2000,0.000000)
        (2001,0.000000)
        (2002,0.000000)
        (2003,1.000000)
        (2004,0.000000)
        (2005,1.000000)
        (2006,5.000000)
        (2007,1.000000)
        (2008,1.000000)
        (2009,9.000000)
        (2010,2.000000)
        (2011,3.000000)
        (2012,3.000000)
        (2013,2.000000)
        (2014,3.000000)
        (2015,2.000000)
        (2016,5.000000)
        (2017,2.000000)
        (2018,3.000000)
        (2019,4.000000)
        (2020,6.000000)
        (2021,6.000000)
        (2022,2.000000)
      };

      \legend{Semantic (223),Loops (96),Destructive (78),Configuration (64)}
    \end{axis}
  \end{tikzpicture}
  \caption{Publication year distribution of all 386 corpus papers, according to the types of software modifications. Papers considering multiple types of modifications are counted multiple times.}
  \label{figure:years_and_approach}

%% file: figure/benchmark_sets.tex
  \centering
  \begin{tikzpicture}
    \begin{axis}[
        width=13em,
        xbar,
        xmin=0,
        xlabel={Number of papers},
        ylabel={Benchmark suite},
        ytick={0,1,2,3,4,5,6,7,8,9,10,11,12},
        yticklabels={Perfect Club,Parboil,{SPEC (Java)},{SPLASH, SPLASH-2},PolyBench,DaCapo,{MediaBench (I, II)},PARSEC,NAS,Rodinia,MiBench,{SPEC (C/C\texttt{++}/Fortran)}},
        yticklabel style={font=\small},
        nodes near coords={\pgfkeys{/pgf/fpu}\pgfmathparse{\pgfplotspointmeta}\pgfmathprintnumber[fixed,precision=0]{\pgfmathresult}},
        nodes near coords align={horizontal},
        bar shift=0pt,
        enlarge x limits={upper, value=0.25},
        enlarge y limits={true, abs value=0.8},
        bar width=0.6em,
        y=0.9em,
        thick,
      ]
      \addplot coordinates {
        (5,0) 
        (6,1) 
        (7,2) 
        (8,3) 
        (11,4) 
        (11,5) 
        (13,6) 
        (13,7) 
        (13,8) 
        (14,9) 
        (14,10) 
        (54,11) 
      };
    \end{axis}
  \end{tikzpicture}
  \caption{Most frequent software benchmark sets (${\ge 5}$) across all 386 corpus papers.}
  \label{figure:benchmark_sets}


%% file: figure/survey_years_and_softset.tex
  \centering
  \begin{tikzpicture}
    \begin{axis}[
        width=20em,
        height=16em,
        ybar stacked,
        ymin=0,
        ymax=35,
        xlabel={Publication year},
        ylabel={Cumulative corpus paper},
        grid,
        bar width=4pt,
        thick,
        cycle list/Set1,
        every axis plot/.append style={fill,fill opacity=0.5},
        x tick label style={/pgf/number format/1000 sep=},
        legend pos=north west,
        reverse legend,
        legend columns=3,
        legend style={nodes={scale=0.8, transform shape}},
      ]
      \addplot coordinates {
        (1992,0.000000)
        (1994,2.000000)
        (1996,0.000000)
        (1997,1.000000)
        (1998,1.000000)
        (1999,2.000000)
        (2000,3.000000)
        (2001,3.000000)
        (2002,0.000000)
        (2003,3.000000)
        (2004,2.000000)
        (2005,2.000000)
        (2006,3.000000)
        (2007,3.000000)
        (2008,3.000000)
        (2009,3.000000)
        (2010,3.000000)
        (2011,3.000000)
        (2012,1.000000)
        (2013,1.000000)
        (2014,2.000000)
        (2015,0.000000)
        (2016,3.000000)
        (2017,0.000000)
        (2018,0.000000)
        (2019,7.000000)
        (2020,1.000000)
        (2021,2.000000)
        (2022,0.000000)
      };
      \addplot coordinates {
        (1992,0.000000)
        (1994,0.000000)
        (1996,0.000000)
        (1997,0.000000)
        (1998,0.000000)
        (1999,0.000000)
        (2000,0.000000)
        (2001,0.000000)
        (2002,0.000000)
        (2003,0.000000)
        (2004,0.000000)
        (2005,0.000000)
        (2006,0.000000)
        (2007,0.000000)
        (2008,0.000000)
        (2009,0.000000)
        (2010,0.000000)
        (2011,1.000000)
        (2012,0.000000)
        (2013,0.000000)
        (2014,2.000000)
        (2015,0.000000)
        (2016,0.000000)
        (2017,0.000000)
        (2018,3.000000)
        (2019,3.000000)
        (2020,3.000000)
        (2021,1.000000)
        (2022,1.000000)
      };
      \addplot coordinates {
        (1992,0.000000)
        (1994,0.000000)
        (1996,0.000000)
        (1997,0.000000)
        (1998,0.000000)
        (1999,0.000000)
        (2000,0.000000)
        (2001,0.000000)
        (2002,0.000000)
        (2003,0.000000)
        (2004,0.000000)
        (2005,0.000000)
        (2006,2.000000)
        (2007,2.000000)
        (2008,1.000000)
        (2009,2.000000)
        (2010,0.000000)
        (2011,0.000000)
        (2012,2.000000)
        (2013,0.000000)
        (2014,0.000000)
        (2015,0.000000)
        (2016,1.000000)
        (2017,0.000000)
        (2018,1.000000)
        (2019,1.000000)
        (2020,0.000000)
        (2021,2.000000)
        (2022,0.000000)
      };
      \addplot coordinates {
        (1992,0.000000)
        (1994,0.000000)
        (1996,0.000000)
        (1997,0.000000)
        (1998,0.000000)
        (1999,0.000000)
        (2000,0.000000)
        (2001,0.000000)
        (2002,0.000000)
        (2003,0.000000)
        (2004,0.000000)
        (2005,0.000000)
        (2006,0.000000)
        (2007,0.000000)
        (2008,0.000000)
        (2009,0.000000)
        (2010,0.000000)
        (2011,0.000000)
        (2012,2.000000)
        (2013,1.000000)
        (2014,1.000000)
        (2015,1.000000)
        (2016,2.000000)
        (2017,1.000000)
        (2018,0.000000)
        (2019,3.000000)
        (2020,1.000000)
        (2021,1.000000)
        (2022,0.000000)
      };
      \addplot coordinates {
        (1992,0.000000)
        (1994,0.000000)
        (1996,0.000000)
        (1997,0.000000)
        (1998,0.000000)
        (1999,0.000000)
        (2000,2.000000)
        (2001,1.000000)
        (2002,1.000000)
        (2003,0.000000)
        (2004,0.000000)
        (2005,0.000000)
        (2006,0.000000)
        (2007,1.000000)
        (2008,1.000000)
        (2009,1.000000)
        (2010,1.000000)
        (2011,1.000000)
        (2012,1.000000)
        (2013,1.000000)
        (2014,0.000000)
        (2015,1.000000)
        (2016,0.000000)
        (2017,0.000000)
        (2018,1.000000)
        (2019,0.000000)
        (2020,0.000000)
        (2021,0.000000)
        (2022,0.000000)
      };
      \addplot coordinates {
        (1992,0.000000)
        (1994,0.000000)
        (1996,0.000000)
        (1997,0.000000)
        (1998,0.000000)
        (1999,0.000000)
        (2000,0.000000)
        (2001,0.000000)
        (2002,0.000000)
        (2003,0.000000)
        (2004,0.000000)
        (2005,0.000000)
        (2006,0.000000)
        (2007,0.000000)
        (2008,0.000000)
        (2009,0.000000)
        (2010,1.000000)
        (2011,2.000000)
        (2012,0.000000)
        (2013,1.000000)
        (2014,1.000000)
        (2015,0.000000)
        (2016,1.000000)
        (2017,1.000000)
        (2018,2.000000)
        (2019,1.000000)
        (2020,0.000000)
        (2021,1.000000)
        (2022,0.000000)
      };
      \addplot coordinates {
        (1992,0.000000)
        (1994,0.000000)
        (1996,0.000000)
        (1997,0.000000)
        (1998,0.000000)
        (1999,0.000000)
        (2000,0.000000)
        (2001,0.000000)
        (2002,0.000000)
        (2003,0.000000)
        (2004,0.000000)
        (2005,0.000000)
        (2006,0.000000)
        (2007,0.000000)
        (2008,0.000000)
        (2009,0.000000)
        (2010,0.000000)
        (2011,0.000000)
        (2012,0.000000)
        (2013,2.000000)
        (2014,0.000000)
        (2015,1.000000)
        (2016,1.000000)
        (2017,0.000000)
        (2018,2.000000)
        (2019,1.000000)
        (2020,1.000000)
        (2021,3.000000)
        (2022,0.000000)
      };
      \addplot coordinates {
        (1992,0.000000)
        (1994,0.000000)
        (1996,0.000000)
        (1997,0.000000)
        (1998,0.000000)
        (1999,0.000000)
        (2000,0.000000)
        (2001,2.000000)
        (2002,1.000000)
        (2003,1.000000)
        (2004,1.000000)
        (2005,1.000000)
        (2006,1.000000)
        (2007,0.000000)
        (2008,2.000000)
        (2009,1.000000)
        (2010,0.000000)
        (2011,0.000000)
        (2012,1.000000)
        (2013,0.000000)
        (2014,0.000000)
        (2015,0.000000)
        (2016,0.000000)
        (2017,0.000000)
        (2018,0.000000)
        (2019,0.000000)
        (2020,0.000000)
        (2021,0.000000)
        (2022,0.000000)
      };
      \addplot coordinates {
        (1992,1.000000)
        (1994,2.000000)
        (1996,1.000000)
        (1997,0.000000)
        (1998,0.000000)
        (1999,3.000000)
        (2000,5.000000)
        (2001,10.000000)
        (2002,1.000000)
        (2003,0.000000)
        (2004,5.000000)
        (2005,1.000000)
        (2006,1.000000)
        (2007,4.000000)
        (2008,2.000000)
        (2009,1.000000)
        (2010,7.000000)
        (2011,4.000000)
        (2012,3.000000)
        (2013,3.000000)
        (2014,12.000000)
        (2015,3.000000)
        (2016,8.000000)
        (2017,1.000000)
        (2018,7.000000)
        (2019,7.000000)
        (2020,7.000000)
        (2021,7.000000)
        (2022,1.000000)
      };

      \legend{SPEC,Rodinia,MiBench,PARSEC,NAS,DaCapo,PolyBench,MediaBench,Other}
    \end{axis}
  \end{tikzpicture}
  \caption{Publication year of all 157 corpus papers that reuse an existing benchmark set.}
  \label{figure:year_and_benchmark}

%% file: figure/benchmark_software.tex
\begin{figure}
  \centering
  \begin{tikzpicture}
    \begin{axis}[
        width=20em,
        xbar,
        xmin=0,
        xlabel={Number of papers},
        ylabel={Benchmark software},
        ytick={0,1,2,3,4,5,6,7,8,9,10,11,12,13,14,15,16,17,18,19,20,21,22,23,24,25,26,27,28,29,30,31},
        yticklabels={{SP  (--/LoneStar/NAS/Rodinia),perlbench  (SPEC),omnetpp  (SPEC),mgrid  (NAS/SPEC),libquantum  (SPEC),h264ref  (SPEC),Blackscholes  (--/CUDA SDK/PARSEC),bitcount  (--/cBench/MiBench),astar  (--/SPEC),vpr  (SPEC),twolf  (SPEC),Tomcatv  (--/SPEC),swim  (SPEC),sphinx3  (SPEC),sjeng  (SPEC),povray  (SPEC),parser  (SPEC),milc  (SPEC),hmmer  (SPEC),compress  (--/SPEC/SPECjvm),SHA  (--/cBench/CHStone/MiBench/Perl Oasis/RiCeps),adi  (--/Livermore Loops/PolyBench/SPEC),LU  (--/NAS/PolyBench/SPLASH-2),ADPCM  (--/cBench/CHStone/MediaBench/PARSEC),JPEG  (--/cBench/MediaBench/MiBench),LBM  (--/Parboil/SPEC),FFT  (--/DSPstone/MiBench/Perfect Club/SPLASH-2),gzip  (--/Chisel/SIR/SPEC),gcc  (--/SPEC),mcf  (SPEC),bzip2  (--/cBench/Chisel/MiBench/SPEC)}},
        yticklabel style={font=\small},
        nodes near coords={\pgfkeys{/pgf/fpu}\pgfmathparse{\pgfplotspointmeta}\pgfmathprintnumber[fixed,precision=0]{\pgfmathresult}},
        nodes near coords align={horizontal},
        bar shift=0pt,
        enlarge x limits={upper, value=0.15},
        enlarge y limits={true, abs value=0.8},
        bar width=0.6em,
        y=0.8em,
        thick,
      ]
      \addplot coordinates {
        (10,0) 
        (10,1) 
        (10,2) 
        (10,3) 
        (10,4) 
        (10,5) 
        (10,6) 
        (10,7) 
        (10,8) 
        (11,9) 
        (11,10) 
        (11,11) 
        (11,12) 
        (11,13) 
        (11,14) 
        (11,15) 
        (11,16) 
        (11,17) 
        (11,18) 
        (11,19) 
        (12,20) 
        (12,21) 
        (15,22) 
        (15,23) 
        (16,24) 
        (17,25) 
        (17,26) 
        (18,27) 
        (19,28) 
        (22,29) 
        (33,30) 
      };
    \end{axis}
  \end{tikzpicture}
  \caption{Software most often targeted ($\ge$ 10 papers)}
  \label{figure:software_names}
\end{figure}

%% file: figure/survey_years_and_language.tex
  \centering
  \begin{tikzpicture}
    \begin{axis}[
        width=20em,
        height=16em,
        ybar stacked,
        ymin=0,
        xlabel={Publication year},
        ylabel={Cumulative corpus paper},
        grid,
        bar width=3pt,
        thick,
        cycle list/Set1,
        every axis plot/.append style={fill,fill opacity=0.5},
        x tick label style={/pgf/number format/1000 sep=},
        try min ticks=5,
        legend pos=north west,
        reverse legend,
      ]
      \addplot coordinates {
        (1978,0.000000)
        (1979,0.000000)
        (1983,1.000000)
        (1984,0.000000)
        (1988,0.000000)
        (1989,1.000000)
        (1992,1.000000)
        (1993,1.000000)
        (1994,1.000000)
        (1996,1.000000)
        (1997,1.000000)
        (1998,3.000000)
        (1999,4.000000)
        (2000,4.000000)
        (2001,6.000000)
        (2002,3.000000)
        (2003,7.000000)
        (2004,6.000000)
        (2005,1.000000)
        (2006,7.000000)
        (2007,5.000000)
        (2008,5.000000)
        (2009,13.000000)
        (2010,6.000000)
        (2011,8.000000)
        (2012,7.000000)
        (2013,9.000000)
        (2014,12.000000)
        (2015,12.000000)
        (2016,14.000000)
        (2017,3.000000)
        (2018,16.000000)
        (2019,21.000000)
        (2020,16.000000)
        (2021,10.000000)
        (2022,4.000000)
      };
      \addplot coordinates {
        (1978,0.000000)
        (1979,0.000000)
        (1983,0.000000)
        (1984,0.000000)
        (1988,0.000000)
        (1989,0.000000)
        (1992,0.000000)
        (1993,0.000000)
        (1994,0.000000)
        (1996,0.000000)
        (1997,0.000000)
        (1998,0.000000)
        (1999,0.000000)
        (2000,0.000000)
        (2001,0.000000)
        (2002,1.000000)
        (2003,0.000000)
        (2004,0.000000)
        (2005,2.000000)
        (2006,1.000000)
        (2007,0.000000)
        (2008,0.000000)
        (2009,0.000000)
        (2010,2.000000)
        (2011,2.000000)
        (2012,2.000000)
        (2013,5.000000)
        (2014,6.000000)
        (2015,6.000000)
        (2016,5.000000)
        (2017,7.000000)
        (2018,6.000000)
        (2019,4.000000)
        (2020,5.000000)
        (2021,6.000000)
        (2022,1.000000)
      };
      \addplot coordinates {
        (1978,1.000000)
        (1979,0.000000)
        (1983,0.000000)
        (1984,1.000000)
        (1988,0.000000)
        (1989,0.000000)
        (1992,0.000000)
        (1993,0.000000)
        (1994,0.000000)
        (1996,0.000000)
        (1997,0.000000)
        (1998,0.000000)
        (1999,1.000000)
        (2000,2.000000)
        (2001,3.000000)
        (2002,2.000000)
        (2003,3.000000)
        (2004,1.000000)
        (2005,0.000000)
        (2006,3.000000)
        (2007,3.000000)
        (2008,0.000000)
        (2009,1.000000)
        (2010,1.000000)
        (2011,2.000000)
        (2012,1.000000)
        (2013,0.000000)
        (2014,0.000000)
        (2015,1.000000)
        (2016,3.000000)
        (2017,2.000000)
        (2018,3.000000)
        (2019,4.000000)
        (2020,6.000000)
        (2021,2.000000)
        (2022,0.000000)
      };
      \addplot coordinates {
        (1978,0.000000)
        (1979,0.000000)
        (1983,0.000000)
        (1984,0.000000)
        (1988,0.000000)
        (1989,0.000000)
        (1992,0.000000)
        (1993,0.000000)
        (1994,0.000000)
        (1996,0.000000)
        (1997,0.000000)
        (1998,0.000000)
        (1999,0.000000)
        (2000,0.000000)
        (2001,0.000000)
        (2002,0.000000)
        (2003,0.000000)
        (2004,0.000000)
        (2005,0.000000)
        (2006,0.000000)
        (2007,0.000000)
        (2008,2.000000)
        (2009,0.000000)
        (2010,3.000000)
        (2011,1.000000)
        (2012,2.000000)
        (2013,0.000000)
        (2014,4.000000)
        (2015,2.000000)
        (2016,1.000000)
        (2017,1.000000)
        (2018,3.000000)
        (2019,3.000000)
        (2020,6.000000)
        (2021,2.000000)
        (2022,1.000000)
      };
      \addplot coordinates {
        (1978,0.000000)
        (1979,0.000000)
        (1983,0.000000)
        (1984,0.000000)
        (1988,0.000000)
        (1989,0.000000)
        (1992,0.000000)
        (1993,0.000000)
        (1994,2.000000)
        (1996,0.000000)
        (1997,1.000000)
        (1998,2.000000)
        (1999,1.000000)
        (2000,2.000000)
        (2001,3.000000)
        (2002,1.000000)
        (2003,0.000000)
        (2004,2.000000)
        (2005,1.000000)
        (2006,1.000000)
        (2007,2.000000)
        (2008,0.000000)
        (2009,1.000000)
        (2010,0.000000)
        (2011,0.000000)
        (2012,0.000000)
        (2013,0.000000)
        (2014,0.000000)
        (2015,0.000000)
        (2016,1.000000)
        (2017,0.000000)
        (2018,0.000000)
        (2019,0.000000)
        (2020,1.000000)
        (2021,0.000000)
        (2022,0.000000)
      };
      \addplot coordinates {
        (1978,1.000000)
        (1979,1.000000)
        (1983,0.000000)
        (1984,0.000000)
        (1988,2.000000)
        (1989,0.000000)
        (1992,0.000000)
        (1993,0.000000)
        (1994,0.000000)
        (1996,0.000000)
        (1997,1.000000)
        (1998,0.000000)
        (1999,1.000000)
        (2000,0.000000)
        (2001,0.000000)
        (2002,1.000000)
        (2003,0.000000)
        (2004,1.000000)
        (2005,0.000000)
        (2006,0.000000)
        (2007,0.000000)
        (2008,1.000000)
        (2009,1.000000)
        (2010,2.000000)
        (2011,0.000000)
        (2012,2.000000)
        (2013,0.000000)
        (2014,4.000000)
        (2015,2.000000)
        (2016,6.000000)
        (2017,4.000000)
        (2018,4.000000)
        (2019,5.000000)
        (2020,2.000000)
        (2021,3.000000)
        (2022,0.000000)
      };

      \legend{C/C++ (209),Java (61),Unknown (46),GPU (31),Fortran (21),Other (44)}
    \end{axis}
  \end{tikzpicture}
  \caption{Publication year distribution of all 386 corpus papers, according to the programming language of the software targeted. Papers considering more than one programming language are counted multiple times.}
  \label{figure:year_and_language}

%% file: figure/survey_years_and_soft_size.tex
  \centering
  \begin{tikzpicture}
    \begin{axis}[
        width=20em,
        height=16em,
        ybar stacked,
        ymin=0,
        ymax=1,
        xlabel={Publication year},
        ylabel={Ratio of corpus paper},
        grid,
        bar width=3pt,
        thick,
        cycle list/Set1,
        every axis plot/.append style={draw,fill,fill opacity=0.5},
        x tick label style={/pgf/number format/1000 sep=},
        legend pos=north west,
        reverse legend,
      ]
      \addplot coordinates {
        (1977,NaN)
        (1978,0.000000)
        (1979,0.000000)
        (1980,NaN)
        (1981,NaN)
        (1982,NaN)
        (1983,0.000000)
        (1984,0.500000)
        (1985,NaN)
        (1986,NaN)
        (1987,NaN)
        (1988,0.000000)
        (1989,0.000000)
        (1990,NaN)
        (1991,NaN)
        (1992,0.000000)
        (1993,0.000000)
        (1994,0.625000)
        (1995,NaN)
        (1996,1.000000)
        (1997,0.500000)
        (1998,0.500000)
        (1999,0.333333)
        (2000,0.375000)
        (2001,0.450000)
        (2002,0.375000)
        (2003,0.400000)
        (2004,0.333333)
        (2005,0.062500)
        (2006,0.393939)
        (2007,0.722222)
        (2008,0.437500)
        (2009,0.447917)
        (2010,0.410714)
        (2011,0.192308)
        (2012,0.380952)
        (2013,0.166667)
        (2014,0.222222)
        (2015,0.326087)
        (2016,0.176282)
        (2017,0.117647)
        (2018,0.158046)
        (2019,0.221429)
        (2020,0.310606)
        (2021,0.376984)
        (2022,0.200000)
      };
      \addplot coordinates {
        (1977,NaN)
        (1978,0.000000)
        (1979,0.000000)
        (1980,NaN)
        (1981,NaN)
        (1982,NaN)
        (1983,0.000000)
        (1984,0.000000)
        (1985,NaN)
        (1986,NaN)
        (1987,NaN)
        (1988,0.000000)
        (1989,0.000000)
        (1990,NaN)
        (1991,NaN)
        (1992,0.000000)
        (1993,1.000000)
        (1994,0.125000)
        (1995,NaN)
        (1996,0.000000)
        (1997,0.166667)
        (1998,0.000000)
        (1999,0.000000)
        (2000,0.041667)
        (2001,0.150000)
        (2002,0.125000)
        (2003,0.300000)
        (2004,0.037037)
        (2005,0.437500)
        (2006,0.212121)
        (2007,0.055556)
        (2008,0.166667)
        (2009,0.093750)
        (2010,0.160714)
        (2011,0.256410)
        (2012,0.228571)
        (2013,0.273810)
        (2014,0.263889)
        (2015,0.347826)
        (2016,0.535256)
        (2017,0.666667)
        (2018,0.320115)
        (2019,0.330952)
        (2020,0.260101)
        (2021,0.273810)
        (2022,0.566667)
      };
      \addplot coordinates {
        (1977,NaN)
        (1978,0.000000)
        (1979,0.000000)
        (1980,NaN)
        (1981,NaN)
        (1982,NaN)
        (1983,0.000000)
        (1984,0.000000)
        (1985,NaN)
        (1986,NaN)
        (1987,NaN)
        (1988,0.000000)
        (1989,0.000000)
        (1990,NaN)
        (1991,NaN)
        (1992,0.000000)
        (1993,0.000000)
        (1994,0.125000)
        (1995,NaN)
        (1996,0.000000)
        (1997,0.333333)
        (1998,0.250000)
        (1999,0.208333)
        (2000,0.041667)
        (2001,0.100000)
        (2002,0.000000)
        (2003,0.100000)
        (2004,0.148148)
        (2005,0.187500)
        (2006,0.196970)
        (2007,0.000000)
        (2008,0.104167)
        (2009,0.104167)
        (2010,0.053571)
        (2011,0.217949)
        (2012,0.109524)
        (2013,0.166667)
        (2014,0.222222)
        (2015,0.108696)
        (2016,0.092949)
        (2017,0.049020)
        (2018,0.078736)
        (2019,0.221429)
        (2020,0.169192)
        (2021,0.226190)
        (2022,0.166667)
      };
      \addplot coordinates {
        (1977,NaN)
        (1978,0.000000)
        (1979,0.000000)
        (1980,NaN)
        (1981,NaN)
        (1982,NaN)
        (1983,0.500000)
        (1984,0.000000)
        (1985,NaN)
        (1986,NaN)
        (1987,NaN)
        (1988,0.500000)
        (1989,1.000000)
        (1990,NaN)
        (1991,NaN)
        (1992,0.000000)
        (1993,0.000000)
        (1994,0.000000)
        (1995,NaN)
        (1996,0.000000)
        (1997,0.000000)
        (1998,0.250000)
        (1999,0.125000)
        (2000,0.166667)
        (2001,0.100000)
        (2002,0.125000)
        (2003,0.050000)
        (2004,0.425926)
        (2005,0.187500)
        (2006,0.121212)
        (2007,0.055556)
        (2008,0.041667)
        (2009,0.229167)
        (2010,0.017857)
        (2011,0.102564)
        (2012,0.180952)
        (2013,0.035714)
        (2014,0.041667)
        (2015,0.086957)
        (2016,0.022436)
        (2017,0.019608)
        (2018,0.044253)
        (2019,0.059524)
        (2020,0.108586)
        (2021,0.051587)
        (2022,0.066667)
      };
      \addplot coordinates {
        (1977,NaN)
        (1978,1.000000)
        (1979,1.000000)
        (1980,NaN)
        (1981,NaN)
        (1982,NaN)
        (1983,0.500000)
        (1984,0.000000)
        (1985,NaN)
        (1986,NaN)
        (1987,NaN)
        (1988,0.500000)
        (1989,0.000000)
        (1990,NaN)
        (1991,NaN)
        (1992,0.000000)
        (1993,0.000000)
        (1994,0.000000)
        (1995,NaN)
        (1996,0.000000)
        (1997,0.000000)
        (1998,0.000000)
        (1999,0.125000)
        (2000,0.125000)
        (2001,0.100000)
        (2002,0.375000)
        (2003,0.050000)
        (2004,0.055556)
        (2005,0.125000)
        (2006,0.075758)
        (2007,0.055556)
        (2008,0.125000)
        (2009,0.125000)
        (2010,0.000000)
        (2011,0.230769)
        (2012,0.050000)
        (2013,0.285714)
        (2014,0.125000)
        (2015,0.108696)
        (2016,0.076923)
        (2017,0.088235)
        (2018,0.242529)
        (2019,0.095238)
        (2020,0.090909)
        (2021,0.000000)
        (2022,0.000000)
      };
      \addplot coordinates {
        (1977,NaN)
        (1978,0.000000)
        (1979,0.000000)
        (1980,NaN)
        (1981,NaN)
        (1982,NaN)
        (1983,0.000000)
        (1984,0.500000)
        (1985,NaN)
        (1986,NaN)
        (1987,NaN)
        (1988,0.000000)
        (1989,0.000000)
        (1990,NaN)
        (1991,NaN)
        (1992,1.000000)
        (1993,0.000000)
        (1994,0.125000)
        (1995,NaN)
        (1996,0.000000)
        (1997,0.000000)
        (1998,0.000000)
        (1999,0.208333)
        (2000,0.250000)
        (2001,0.100000)
        (2002,0.000000)
        (2003,0.100000)
        (2004,0.000000)
        (2005,0.000000)
        (2006,0.000000)
        (2007,0.111111)
        (2008,0.125000)
        (2009,0.000000)
        (2010,0.357143)
        (2011,0.000000)
        (2012,0.050000)
        (2013,0.071429)
        (2014,0.125000)
        (2015,0.021739)
        (2016,0.096154)
        (2017,0.058824)
        (2018,0.156322)
        (2019,0.071429)
        (2020,0.060606)
        (2021,0.071429)
        (2022,0.000000)
      };

      \legend{Unknown,Application,Project,File,Function,Block}
    \end{axis}
  \end{tikzpicture}
  \caption{Publication year distribution of all 386 corpus papers, according to the ratio of size of software targeted.}
  \label{figure:year_and_size}

%% file: split_discussion.tex
\section{Discussion}
\label{section:discussion}

In \autoref{section:survey} we examined the results of our survey by examining the identified papers and the characteristics of the non-functional properties they improve, the type of approaches they use, and the benchmark they target.
In this section, we first take a step back and discuss gaps in the literature with regard to real-world needs, the main characteristic to consider when selecting target software, and finally then we give more practical recommendations.

\subsection{Missing literature}

There are obvious discrepancies between real-world software engineering and the papers identified in our survey.
In many aspects, our survey fails to reflect that richness of software development.

First, despite 57\% of the papers targeting real-world software, C/\cpp software is vastly over-represented (54\%), as is Unix/Linux software.
According to the TIOBE index\footnote{\url{https://www.tiobe.com/tiobe-index}}---an indicator of online popularity---Python has recently become the most popular programming language.
In contrast, only four papers in our survey (1\%) consider Python software, with three also coincidentally considering C or \cpp software.

The 2021 Stack Overflow’s annual developer survey\footnote{\url{https://insights.stackoverflow.com/survey/2021}} provides more practical insight regarding developers demographics.
In particular, JavaScript (69\% of respondents), SQL (56\%), Python (42\%), TypeScript (36\%), Node.js (36\%), \csharp (30\%), or Bash/Shell (28\%) are all technologies very popular in industry that are apparently completely missing from academic research.
Likewise, most professional developers use Windows (41\%, whilst 30\% use MacOS and 25\% use a Linux-based distribution), which again is not reflected in our own survey.

Finally, only 6\% of the identified relevant work target software in mobile devices.
In the meantime, in 2022 the number of active Android devices is frequently reported as being as high as 3 billions (1.8 billions for iOS devices).
For these mobile devices, non-functional properties such as memory usage and energy consumption are absolutely critical.

\subsection{Software characteristics}

Representation in terms of the choice of target software is critical.
In this section we highlight five software characteristics that should be considered when considering how techniques that improve non-functional properties of software might generalise.
In general, whilst some opportunities for improvement \emph{might} be universal, it is sensible to think that some can only be detected and specially taken care of when considering a large panel of diverse software.

\begin{description}
\item[Programming language.]
  Programming languages are driven by different coding paradigms and development best practices, have access to entirely different libraries, and follow vastly different syntax.
  The vast majority of published research target C/\cpp or Java software, ignoring the popularity of languages such as Python, JavaScript, PHP, or SQL.
  Work is also almost entirely centred on imperative programming features, neglecting functional functionalities essential, e.g., for languages such as Haskell or Scala.
\item[Size.]
  Software size is the second most indisputable critical characteristic.
  Small programs can be maintained much more rigorously, expose fewer and harder inefficiencies, and overall provide much less material for repairs, which may impede some types of approaches~\cite{barr:2014:fse}.
  On the other hand, large software with hundred or more files may be difficult to improve for the complete opposite reasons, as inefficiencies may be more numerous and potentially simpler to deal with, but also be much harder to locate.
\item[Architecture/Application.]
  There are as many types of software as there are types of applications, and each may expose different specific types of inefficiencies.
  It is important that research is conducted on all types of software, including for example GUI and terminal-based/command-line software, single- and multi-purpose software, model-view-controller and monolithic software, single-core and message-passing software, desktop/mobile/embedded software, or general libraries/APIs and specific applications.
\item[Application.]
  Similarly, the application domain the selected software targets may strongly impact the types of inefficiencies it might expose.
  We can cite for example system software (e.g., kernels, compilers, drivers, general utility tools), media-related software (e.g., compression, image/video), scientific software (e.g., machine learning, genomics), or games.
\item[Number of contributors.]
  Industry practices, as well as large open-source software, involve many contributors that often don't share a clear or deep understanding of the system in its entirety.
  As with other characteristics, software written by a single developer, a small team, multiple teams, or many infrequent contributors, may expose different types of inefficiencies.
\end{description}

\subsection{Software Selection and Experimental Protocol Recommendations}

There is yet no standard benchmark specifically tailored to the improvement of non-functional properties of software.
Whilst SPEC is the closest thing, it was proposed with hardware comparison and compiler optimisation in mind, and is not particularly well suited for potentially destructive evolutionary approaches.
Indeed, SPEC is designed for applications in which semantic changes are not expected.
As such, running the optimised software on few known inputs is sufficient to adequately compare performance.
However, as soon as potential semantic changes are introduced (e.g., through parameterisation, or destructive source code changes), it becomes essential to control for correctness and generalisation.
Hence, benchmarks crafted for general improvement of non-functional properties of software should ideally consider software providing a comprehensive test suite or at least software for which a large amount of input data is available.

In terms of experimental protocol, one should generally follow the example of machine learning research, that provides a variety of strong procedures to ensure that the performance improvements are generalisable and reproducible.
Whilst the simplest isolated holdout method ---in which training and parameterisation should be performed on data disjoint from the data used for actual performance comparison--- may seem reasonable, especially for stochastic methods and methods introducing semantic changes we strongly advocate cross-validation.

%% file: split_threats.tex
\section{Threats to Validity}
\label{section:empirical}


\textbf{Keywords used in the repository search may not cover all relevant literature.}
Due to its restrictive nature, there might be whole types of relevant work that a keyword search is not able to reveal.
To mitigate that threat, we conducted first a preliminary search to discover all potential keywords.
We tried to hand-pick a large number of papers (100) using very diverse traits, including research fields, types of non-functional properties improved, programming language, various synonyms, etc.
We then extracted from their titles and abstracts generic terms and performed a frequency analysis, subsequently used to select the most potentially effective keywords.

\textbf{Not all major publishers have been directly queried.}
Indeed, online libraries such as Springer Link or Science Direct haven't been considered due to their lack of complex query ability.
To mitigate this threat, we considered a healthy combination of primary and secondary sources of work.
First, we choose ACM and IEEE, two of the main publishers in computer science\footnote{\url{https://www.spinellis.gr/blog/20170915/index.html}}, both for conferences and journals.
Then, we considered Scopus, as in addition to being the online library of Elsevier it also indexes many other publishers and in particular Springer, who publishes the Lecture Notes in Computer Science (LNCS) series covering many conference proceedings in all areas of computer science.
Finally, we also considered both Google Scholar and ArXiV to further increase the potential coverage of the survey.

\textbf{The repository search wasn't exhaustive.}
Another threat is the very high number of papers returned by the digital libraries (as clearly shown in \autoref{table:systematic_pass}), despite the restrictive compound queries.
We choose to consider the 200 first papers returned by each query, rather than use even more restrictive queries, to make sure not to miss papers otherwise covered.
We also assume that 200 hundred papers, i.e., for example ten pages down on Google Scholar, is a reasonable limit to what would be investigated manually.
The repository search methodology, adapted from~\cite{hort:2022:tse}, then ensures that such a large number of papers (up to 5000) can be effectively considered.

\textbf{Relevant work may not have been correctly identified.}
The identification of relevant work was a long repetitive manual task, and involuntary errors may impact conclusions drawn from our survey.
The main concern here is the relevance of the paper selection during the survey.
To mitigate this threat, we sampled 100 entries uniformly at random from the 3749 unique papers yielded by the repository search and cross-checked that the resulting relevant work was the same when processed independently by both authors of the survey.

Finally, \textbf{the classification used in \autoref{section:survey}'s survey may not reflect the state-of-the-art.}
In order to validate both our results and our conclusions we turned toward prominent authors of the literature.
In total 1080 authors have been identified throughout our survey (on average 2.8 authors per paper).
More precisely, there are 31 authors of three papers, nine authors of four papers, and seven authors of five or more papers.
All 47 authors thus identified have been contacted and their feedback has been used to improve this survey.

%% file: split_conclusion.tex
\section{Conclusions}
\label{section:conclusion}

We provide a comprehensive survey of empirical work on non-functional property improvement of software.
We also provide a list of benchmarks and recommendations for future work in this domain.
We hope our findings and the artefact will drive more research in the area.